\documentclass[structabstract]{aa}  
\newcommand{\Ms}{$M_{\odot}$}

\newcommand{\rs}{R$_{\star}$}
\newcommand{\env}{$\sim$}
\newcommand{\fors}{Mg$_2$SiO$_4$}
\newcommand{\ens}{MgSiO$_3$}
\newcommand{\spin}{MgAl$_2$O$_4$}
\newcommand{\sili}{SiO$_2$}
\newcommand{\al}{Al$_2$O$_3$}
\newcommand{\tio}{TiO$_2$}

\newcommand{\mic}{$\mu$m}

\newcommand{\kms}{km s$^{-1}$}

\newcommand{\cmc}{cm$^{-3}$}

\usepackage{graphicx}

\usepackage{color}
\usepackage{txfonts}
\begin{document}

\title{Dust formation in the oxygen-rich AGB star IK Tau}

\author{D. Gobrecht\inst{\ref{inst1},\ref{inst2}}\and I. Cherchneff\inst{\ref{inst2}}\thanks{Corresponding author: isabelle.cherchneff@unibas.ch}\and A. Sarangi\inst{\ref{inst3},\ref{inst4}}, J.M.C. Plane\inst{\ref{inst5}} \and S. T. Bromley\inst{\ref{inst6},\ref{inst7}}}
\institute{Osservatorio Astronomico di Teramo, INAF, 64100 Teramo, Italy\label{inst1}
\and
Departement Physik, Universit{\"a}t Basel, Klingelbergstrasse 82, CH-4056 Basel, Switzerland\label{inst2}
\and
NASA Goddard Space Flight Center, Greenbelt, MD 20771, USA\label{inst3}
\and
Physics Department, The Catholic University of America, Washington, DC 20064, USA\label{inst4}
\and
School of Chemistry, University of Leeds, Leeds LS2 9JT, United Kingdom\label{inst5}
\and
Departament de Quimica Fisica, Universitat de Barcelona, Mart' i Franqu{\`e}s 1, E-08028 Barcelona, Spain\label{inst6}
\and
Instituci{\'o} Catalana de Recerca i Estudis Avan{\c c}ats (ICREA), 08010 Barcelona, Spain\label{inst7}}

  \date{Submitted 19 November 2014; Accepted 12 October 2015}
\abstract {} 
   {We model the synthesis of molecules and dust in the inner wind of the oxygen-rich Mira-type star IK Tau by considering the effects of periodic shocks induced by the stellar pulsation on the gas and by following the non-equilibrium chemistry in the shocked gas layers between 1~\rs\ and 10~\rs. We consider a very complete set of molecules and dust clusters, and combine the nucleation phase of dust formation with the condensation of these clusters into dust grains. We also test the impact of increasing the local gas density. Our derived molecular abundances and dust properties are compared to the most recent observational data.}
  {A semi-analytical formalism based on parameterised fluid equations is used to describe the gas density, velocity, and temperature in the inner wind. The chemistry is described by using a chemical kinetic network of reactions and the condensation mechanism is described by a Brownian formalism. A set of stiff, ordinary, coupled differential equations is solved, and molecular abundances, dust cluster abundances, grain size distributions and dust masses are derived.}
   {The shocks drive an active non-equilibrium chemistry in the dust formation zone of IK Tau where the collision destruction of CO in the post-shock gas triggers the formation of C-bearing species such as HCN and CS. Most of the modelled molecular abundances agree well with the latest values derived from Herschel data, except for SO$_2$ and NH$_3$, whose formation may not occur in the inner wind. Clusters of alumina, \al, are produced within 2~\rs\ and lead to a population of alumina grains close to the stellar surface. Clusters of silicates (\fors) form at larger radii ($r > 3$~\rs), where their nucleation is triggered by the formation of HSiO and H$_2$SiO. They efficiently condense and reach their final grain size distribution between $\sim 6$~\rs\ and 8~\rs\ with a major population of medium size grains peaking at $\sim 200$ \AA. This two dust-shell configuration agrees with recent interferometric observations. The derived dust-to-gas mass ratio for IK Tau is in the range $1-6 \times 10^{-3}$ and agrees with values derived from observations of O-rich Mira-type stars. }
   {Our results confirm the importance of periodic shocks in chemically shaping the inner wind of AGB stars and providing gas conditions conducive to the efficient synthesis of molecules and dust by non-equilibrium processes. They indicate that the wind acceleration will possibly develop in the radius range $4-8$~\rs\ in IK Tau.  }


\titlerunning{Dust formation in IK Tau}
\authorrunning{Gobrecht et al.}

\maketitle
\section{Introduction}
Low mass stars with masses between 1~\Ms~and 4~\Ms~ascend the asymptotic giant branch (AGB) in the late stages of their evolution where they experience thermal pulses, dust formation in their cool extended atmosphere and mass loss. AGB stars are the prevalent dust providers to galaxies through their strong winds that replenish the surrounding medium with molecules and dust grains (\cite{ge89}). Depending on the number of dredge-up events experienced by the star and its evolutionary stage on the AGB, the red giant forms either oxygen-rich or carbon-rich solid grains reflecting the carbon-to-oxygen (hereafter C/O) ratio of the stellar photosphere. The astronomical evidence for the synthesis of dust in AGBs is brought by the identification of typical dust features in the spectral energy distribution (SED) of AGB stars, including the 9.7 \mic~and 18 \mic\ features of silicates in oxygen-rich AGBs and the $11.3$ \mic~feature ascribed to silicon carbide grains in carbon stars. Recent surveys with the Herschel space telescope of the Magellanic Clouds have questioned the primacy of AGBs as galactic dust providers at lower metallicity than that of the Milky Way (\cite{mat09}; \cite{boy12}). However, owing to their large number in stellar populations, AGB stars remain important dust producers in the local and high redshift universe. 

Despite their crucial role in the dust budgets of galaxies, the mechanisms responsible for the production of dust in AGB winds are still largely unknown. From interferometric data, we know that the dust formation region is located within \env~10~\rs~from the stellar photosphere (\cite{dan94}). Periodic shocks resulting from the stellar pulsation pervade the region, lift dense and warm gas layers to larger radii, and facilitate the formation of solids and the acceleration of the stellar wind by radiation pressure on dust grains because of their high dust opacity (\cite{bow88}). These dense layers are also conducive to the formation of molecules, and many gas-phase species have been identified close to the stellar surface in both oxygen- and carbon-rich AGBs by various observing facilities, including most recently Herschel (e.g., \cite{cer08}, \cite{gro11}, \cite{scho11}, \cite{jus12}). Furthermore, these layers are characterised by non-equilibrium chemical processes induced by the periodic passage of the shocks that leads to the formation of molecules and dust grains. For molecules, the chemistry in the photosphere is controlled by the C/O ratio and the formation of carbon monoxide, CO. This molecule traps all oxygen (carbon) atoms in a carbon-rich (oxygen-rich) gas. The partial dissociation of CO by the shocks releases oxygen atoms in carbon-rich environments and carbon atoms in oxygen-rich environments, leading to a dual chemistry that forms O-bearing species like water, H$_2$O, in the wind of carbon stars, and C-bearing species in the wind of O-rich AGBs (\cite{wil98}, \cite{dua99}, Cherchneff 2006, 2011a, 2012). The widespread occurrence of warm water in several carbon stars has been highlighted with Herschel (\cite{neu11}) whereas hydrogen cyanide, HCN, was detected close to the stellar surface in O-rich Miras, again with Herschel (\cite{jus12}). These detections confirm the crucial role of shock-induced chemistry in the formation of these unexpected molecules in AGB stellar winds, albeit other processes have been proposed that may also contribute in breaking the chemical dichotomy (\cite{ag10}). 

The SEDs of oxygen-rich AGB stars show spectral bands typical of alumina (\al) in the range $9-15$~\mic, of silicates  around $9.7$ \mic\ and $18$ \mic, or a mix of both. Furthermore, stars that ascend the AGB show trends in dust production (\cite{lit90}). The less evolved, low mass-loss rate stars show the presence of alumina (\al), while more evolved stars characterised by high mass-loss rates show silicate features. Observations point to the synthesis of large grains ($a$~$>$~$0.3$~\mic) close to the star in a few O-rich objects that include semi-regular variable stars (\cite{nor12}), while the presence of alumina and/or silicates close to the star ($r < 2$~\rs) and at larger radius ($r \sim 5$~\rs), respectively, have been inferred to explain recent interferometric data (\cite{witt07, sac13, kar13}). Because the alumina and silicate opacities are low in the wavelength range where most of the stellar flux is emitted, scattering on large grains ($a > 0.1$ \mic) was proposed as a viable mechanism to accelerate the wind in O-rich AGBs at small radii (\cite{hof08, bla12}). 

Despite the astronomical evidence of alumina and silicates in O-rich AGBs, the processes controlling the formation of dust in these stars are not well understood. Previous theoretical studies on dust production in O-rich AGB stars have relied on homogeneous or heterogeneous nucleation theory. They usually assume that chemical equilibrium prevails in the dust formation zone, and derive the gas phase composition from equilibrium conditions. They further assume that the growth of dust grains proceeds through surface deposition of atoms and molecules, and the final size of grains is controlled by accretion and evaporation under equilibrium conditions  (\cite{gail99}, Ferraroti \& Gail 2002, 2006, \cite{woit06}, \cite{hof08}, \cite{della14}). Some investigations have pointed to the fact that silicates and oxides such as alumina do not have gas-phase homogeneous monomers out of which condensation of grains proceeds, and that heterogeneous condensation may occur through the deposition of chemical species on the surface of pre-existing grain seeds of different chemical type (e.g., alumina, \al, or titanium oxide, \tio) (\cite{gail98,jeo03}, Ferraroti \& Gail 2002, 2006). In some models that aim to explain wind acceleration and reproduce the photometric properties of O-rich AGBs, a pre-requisite to dust growth by surface deposition is the assumption that a population of small dust seeds exists (Bladh et al. 2012, 2013, 2015). Neither the formation of these dust grains nor the formation of molecules from the gas phase by non-equilibrium processes is treated in such approaches. A thorough review of these various topics and mechanisms involved in homogeneous and heterogenous nucleation at chemical equilibrium is provided by Gail \& Sedlmayr (2013). Recently, the role of titanate clusters as seed nuclei in AGB winds was investigated (\cite{pla13, gou13}), while Goumans \& Bromley (2012) studied the formation pathways to silicate dust from the gas phase. In this latter study, the growth of enstatite, \ens, and forsterite, \fors, clusters follows the combination of oxygen addition and magnesium inclusion steps. Depending on the gas conditions, these chemical routes become exergonic and lead to the efficient synthesis of enstatite and forsterite dimers. Whatever the chemical processes involved in the synthesis of dust are, it is likely that the dense, shocked layers above the photosphere of AGB stars provide the ideal gas conditions for the synthesis of molecules, dust clusters and the growth of dust grains by non-equilibrium processes (Cherchneff 2012, 2013).

Building on these findings and on a previous study by Duari et al. (1999), we present in this paper a complete chemical description of the inner wind of the oxygen-rich AGB star, IK Tauri (or NML Tauri). This exhaustive model describes the formation of molecules and dust grains in the inner wind of an oxygen-rich AGB star and is based on a non-equilibrium, chemical kinetic approach. We consider the impact of the periodic shocks on the gas layers above the photosphere and study the gas and the solid phase of the extended atmosphere, which corresponds to the dust formation and wind acceleration zone between 1 and 10~\rs. We model the formation of various chemical species, some of which have being recently detected in the inner wind of IK Tau with Herschel, and simultaneously model the formation of oxygen-rich clusters, including silicates and alumina, out of the gas phase. We then follow the condensation of these clusters to form dust grains and derive grain size distributions. This study aims to derive trends on the molecular component and the formation of dust in the inner wind of IK Tau, but the results may equally apply to other O-rich stars in the Mira phase on the thermally-pulsing AGB.

The paper is structured as follows: we briefly present the physical model used to describe the shocked inner wind of IK Tau in \S \ref{sec2}. The chemistry of molecule and cluster formation, and the condensation formalism are described in \S \ref{sec3}. Results on molecules, clusters and dust are presented in \S \ref{sec4}, and finally, a discussion follows in \S \ref{sec5}. 

\begin{table}
\caption{Stellar and inner wind parameters for IK Tau. The wind parameters are explained in \S~\ref{physmod}.}           
\label{tab1}       
\centering                         
\begin{tabular}{l l l}        
\hline\hline      
\multicolumn{3}{c}{Star}\\
\hline          
M$_{\star,ZAMS} $ & 1.5 \Ms &  1 \\   
\rs  & 2.5 $\times 10^{13}$ cm & 2 \\
{C/O} & 0.75 & 3 \\  
P & 470 days & 4 \\
T$_{eff}$ & 2200 K & 2 \\
\hline
\multicolumn{3}{c}{Wind}\\
\hline
r$_s$ & 1 \rs & 3 \\
V$_s$ & 32 \kms & 3 \\
$\dot{{\rm M}}$ & $8 \times 10^{-6}$ \Ms\ yr$^{-1}$& 2\\
v$_{esc}$ & 17.7 \kms\ & 2 \\
$\gamma$ & 0.98 & \\
$ \alpha$ & $-0.6$  & 3\\
n$_{gas}$(r$_s$)  & 3.62 $\times 10^{14}$ \cmc  & 3 \\
T$_{gas}$(r$_s$) & 2200 K & -- \\
\hline                                    
\end{tabular}
\tablebib{
1) Fox \& Wood (1982) ; 2) Decin et al 2010; 3) Duari et al. (1999); 4) Hale et al. (1997)}
\end{table}

\section{Model of the inner wind}
\label{sec2}

\begin{figure*}
\centering
  \includegraphics[width=16.0cm]{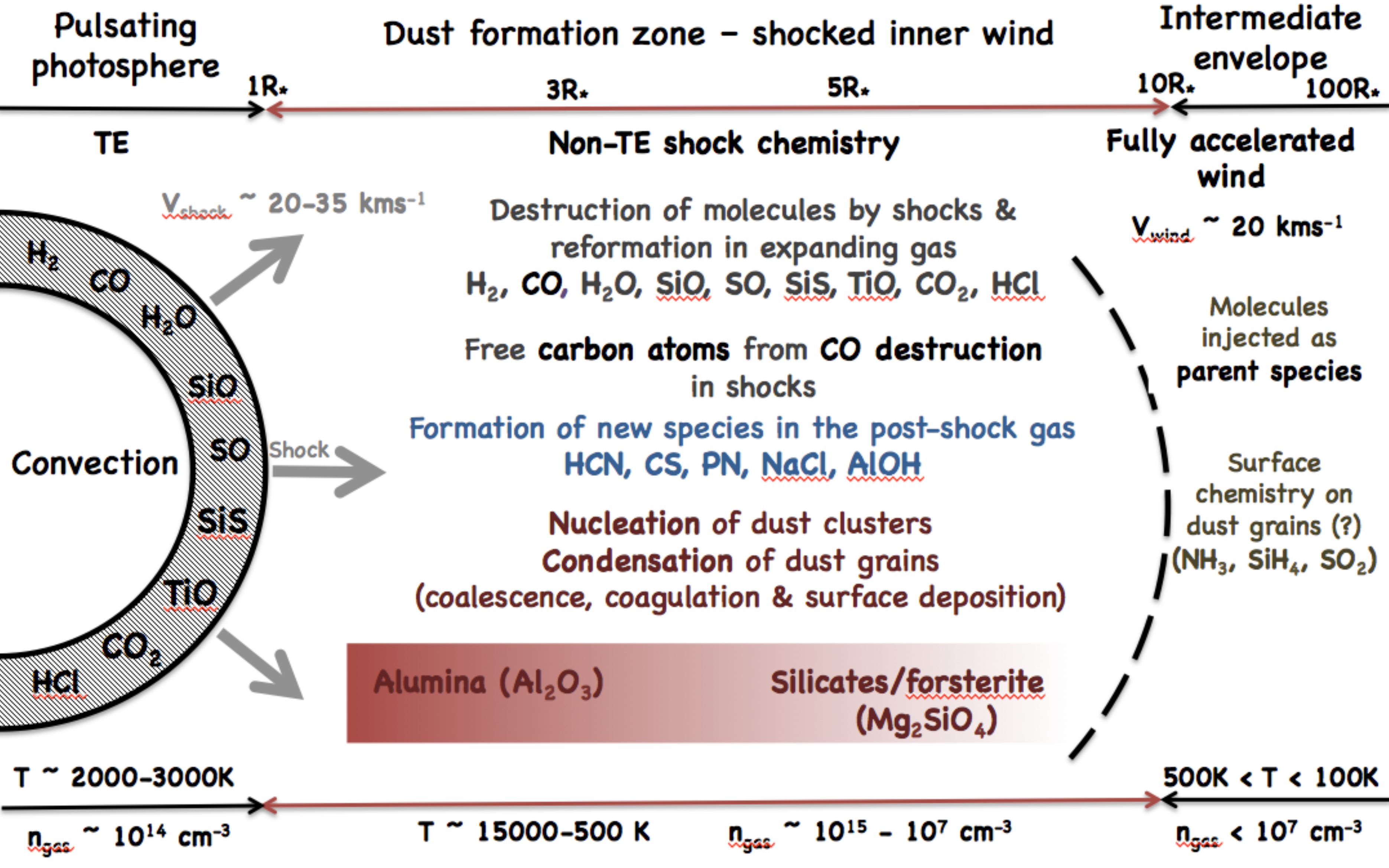}
\caption{Schematic view of the dust formation zone in O-rich AGB stars, which includes typical physical parameters and the prevalent chemical processes related to dust production. Molecules present under TE in the photosphere are shown.   }
\label{fig1}
\end{figure*}

The photosphere is assumed to be at thermodynamic equilibrium (TE) and is characterised by one gas temperature and pressure at which molecular abundances are estimated. We then shock the gas layer at the photosphere, which gradually moves to larger radii where it is again shocked, as described in \S\ \ref{physmod}. We assume spherical symmetry of the dust formation zone, and that one shock propagates in the inner wind at each pulsation period. 
In the chemical description of the gas phase, we consider all molecules detected close to the stellar surface in O-rich Miras stars, and the formation of dust clusters that triggers the condensation of dust grains. We consider two chemical types of dust, silicates and alumina, and the molecular clusters that partake in the formation of small condensation nuclei. We condense the dust seeds as soon as they form, by assuming they coalesce and coagulate to form dust grains. The studied region spans 1~\rs \ to 10~\rs, and is illustrated in Figure \ref{fig1}, where the gas parameters and the main chemical and physical processes are indicated. 

\subsection{Physical model}
\label{physmod}

\begin{table*}
  \caption{Gas parameters including the shock velocity V$_s $, the Mach number $M$, the gas temperature, and the gas number density in the dust formation zone of IK Tau as a function of radius $r$. Numbers in parenthensis give powers of 10. } 
  \label{tab2}
\centering
\begin{tabular}{ l l l  l  l  l  l  l  l    }
 \hline \hline
 & & &\multicolumn{2}{c}{preshock}&\multicolumn{2}{c}{shock front} &\multicolumn{2}{c}{post-shock  $\phi = 0.$} \\
 r & V$_s $ & $M$ & n$_0 $ & T$_0 $ & n$_s $ & T$_s$ & n$_{e} $ & T$_e$  \\
 \hline                       
1.0  &  32.0  & 9.8 & 3.6(14) & 2200 & 2.1(15) & 42824 & 1.2(16) & 6295  \\
1.5  &  26.1  & 9.0 & 7.8(12) & 1725 & 4.4(13) & 28793 & 2.2(15) & 4741  \\
2.0  &  22.6  & 8.5 & 7.2(11) & 1451 & 4.1(12) & 21712 & 1.9(14) & 3858  \\
2.5  &  20.2  & 8.1 & 1.4(11) & 1270 & 7.7(11) & 17473 & 3.4(13) & 3277  \\
3.0  &  18.5  & 7.8 & 3.9(10) & 1138 & 2.2(11) & 14638 & 9.4(12) & 2874  \\
3.5  &  17.1  & 7.6 & 1.5(10) & 1037 & 8.1(10) & 12630 & 3.3(12) & 2568  \\
4.0  &  16.0  & 7.4 & 6.5(9)  & 958  & 3.6(10) & 11098 & 1.4(12) & 2340  \\
5.0  &  14.3  & 7.1 & 1.9(9)  & 838  & 1.0(10) &  8930 & 3.8(11) & 1997  \\
6.0  &  13.1  & 6.8 & 7.2(8)  & 751  & 3.9(9)  &  7497 & 1.3(11) & 1748  \\
7.0  &  12.1  & 6.6 & 3.4(8)  & 684  & 1.8(9)  &  6460 & 6.0(10) & 1574  \\
8.0  &  11.3  & 6.4 & 1.8(8)  & 632  & 9.9(8)  &  5691 & 3.1(10) & 1419  \\
9.0  &  10.7  & 6.3 & 1.1(8)  & 589  & 5.9(8)  &  5082 & 1.8(9)  & 1307  \\
10.0 &  10.1  & 6.2 & 7.1(7)  & 553  & 3.8(8)  &  4597 & 1.1(9)  & 1209  \\
 \hline 
\end{tabular}
\tablefoot{Radius $r$ is in~\rs, shock velocity V$_s$ in \kms, temperatures T$_{0, s, e}$ in Kelvin, and gas number densities n$_{0, s, e}$ in \cmc.}
\end{table*}

To describe the physical conditions experienced by the upper atmosphere of IK Tau, lifted by the recurrent passage of shocks induced by the stellar pulsation, we use the semi-analytical model of the inner wind of AGB stars developed by Cherchneff et al. (1992) and Cherchneff (1996), and used by Willacy \& Cherchneff (1998), Duari et al. (1999), \cite{cau02}, and Cherchneff (2006, 2011, 2012). The post-shock gas consists of 1) a narrow layer in the immediate back of the shock front where the highly endothermic, collision-induced dissociation of molecular hydrogen provides the gas cooling through the reaction ${\rm H}_2~+~{\rm H \rightarrow H + H + H}$, and 2) a region where the gas cools by adiabatic expansion, expands to larger radii, and falls back to its initial position owing to the stellar gravity. 

The narrow layer is modelled by using the study of molecular shocks in AGB stars by Fox \& Wood (1985). We derive the gas parameters in the shock front from the Rankine-Hugoniot jump conditions for a shocked gas, and assume the shock is weakly radiative. The gas cooling is provided by H$_2$ dissociation, which occurs over a space length in the postshock gas given by
\begin{equation}
\label{eq0}
   {\rm l}_{diss} = \tau_{diss} \times {\rm v}_{gas} =  {\frac{1}{{\rm k}_{diss}\times {\rm n(H}_2{\rm)} }} \times \left[{\frac{{\rm V}_s}{{\rm N}_{jump}}}\right],
\end{equation}
where n(H$_2$) is the number density of H$_2$ in the post-shock gas, k$_{diss}$ is the rate of the dissociation reaction of H$_2$, V$_s$ is the shock velocity at the shock formation radius r$_s$, as given in Table \ref{tab1}, and N$_{jump}$ is the Rankine-Hugoniot velocity shock jump. The H$_2$ dissociation cooling length behind the shock takes values that span several orders of magnitude ($1-10^5$ cm) depending on the shock amplitude (\cite{wil98}). 
 
Modelling the adiabatic cooling region is based on the study by \cite{ber85}, who derived analytical solutions to the hydrodynamic fluid equations that describe the state of the gas periodically shocked in the inner wind of a Mira star. The inner wind in IK Tau therefore consists of gas layers that move forwards and fall back to their initial position over a pulsation period, gradually moving outwards owing to shock energy dispersion. For the purposes of this study, we assume that the gas layers follow strictly periodic motions close to the star and fall back to their initial position in the inner wind before the passage of the next shock. 

The initial time-averaged conditions of the pre-shock gas at each radius follows the formalism developed by Cherchneff et al. (1992). For the gas density, the variation as a function of stellar radius depends on the extended scale height and is given by  
\begin{equation}
\label{eq01}
\rho(r) = \rho(r_s)\times \exp\Bigg( -{r_s(1-\gamma^2) \over{H_0(r_s)(1-\alpha)}} \Big[1- \Big( {r \over{r_s}} \Big)^{(\alpha-1)} \Big] \Bigg), 
\end{equation}
where $H_0(r_s)$ is the static scale height at radius $r_s$, and the parameter $\gamma$ is defined as the ratio of the shock velocity and the local escape velocity ($\gamma = V_s/v_e(r_s)$). The parameter $\alpha$ is the exponent of the power law followed by the gas temperature given by 
\begin{equation}
\label{eq02}
T(r) = T(r_s) \times \Big( {r \over r_s} \Big)^{\alpha},
\end{equation}
where $\alpha=-0.6$ (\cite{dua99}). Equation \ref{eq01} accounts for the effect of the shocks on the local gas density (\cite{cher92}).

\begin{figure}
 \centering
 \includegraphics[width=\columnwidth]{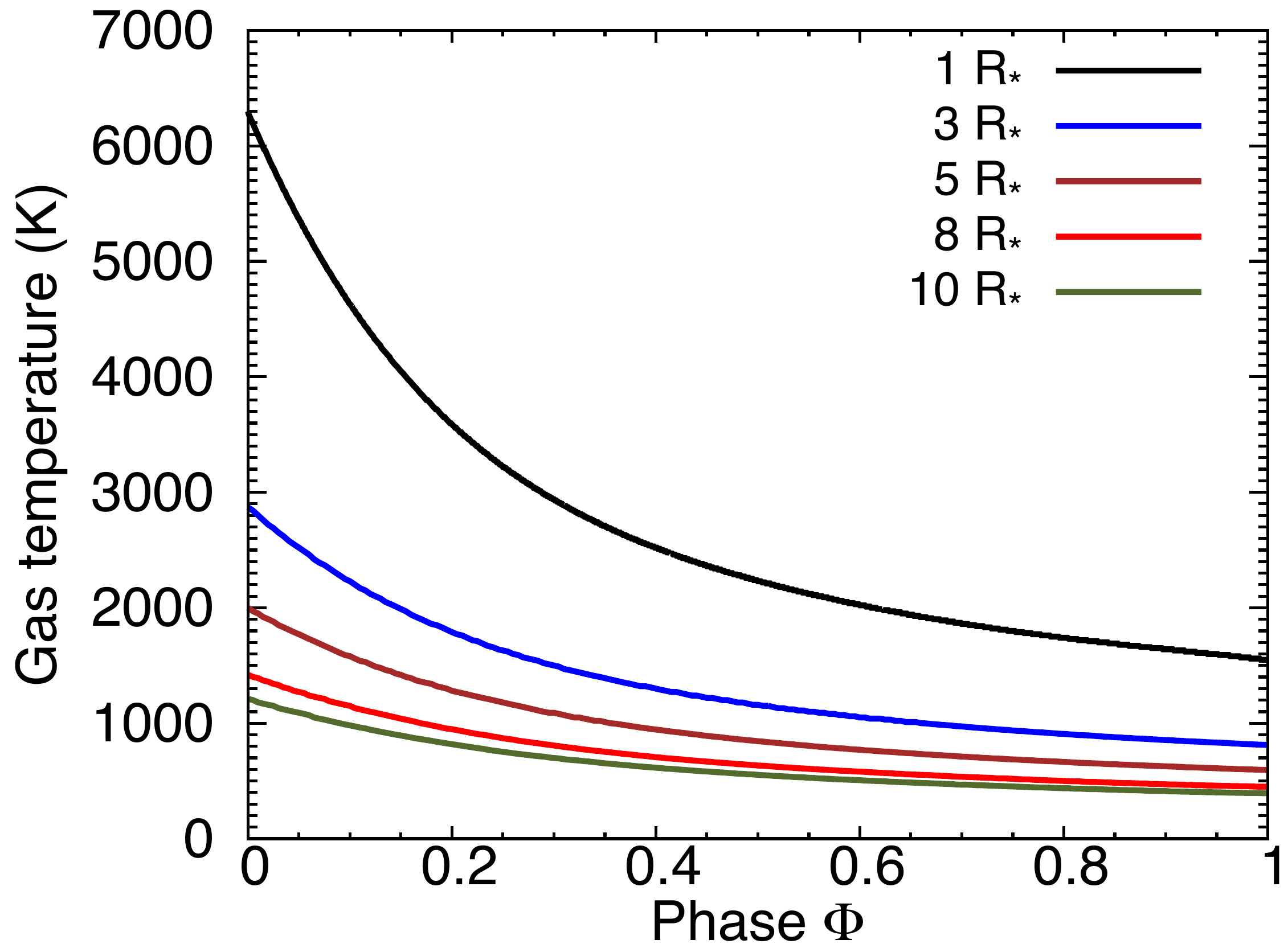} 
  \includegraphics[width=9.1cm,height=6.7cm]{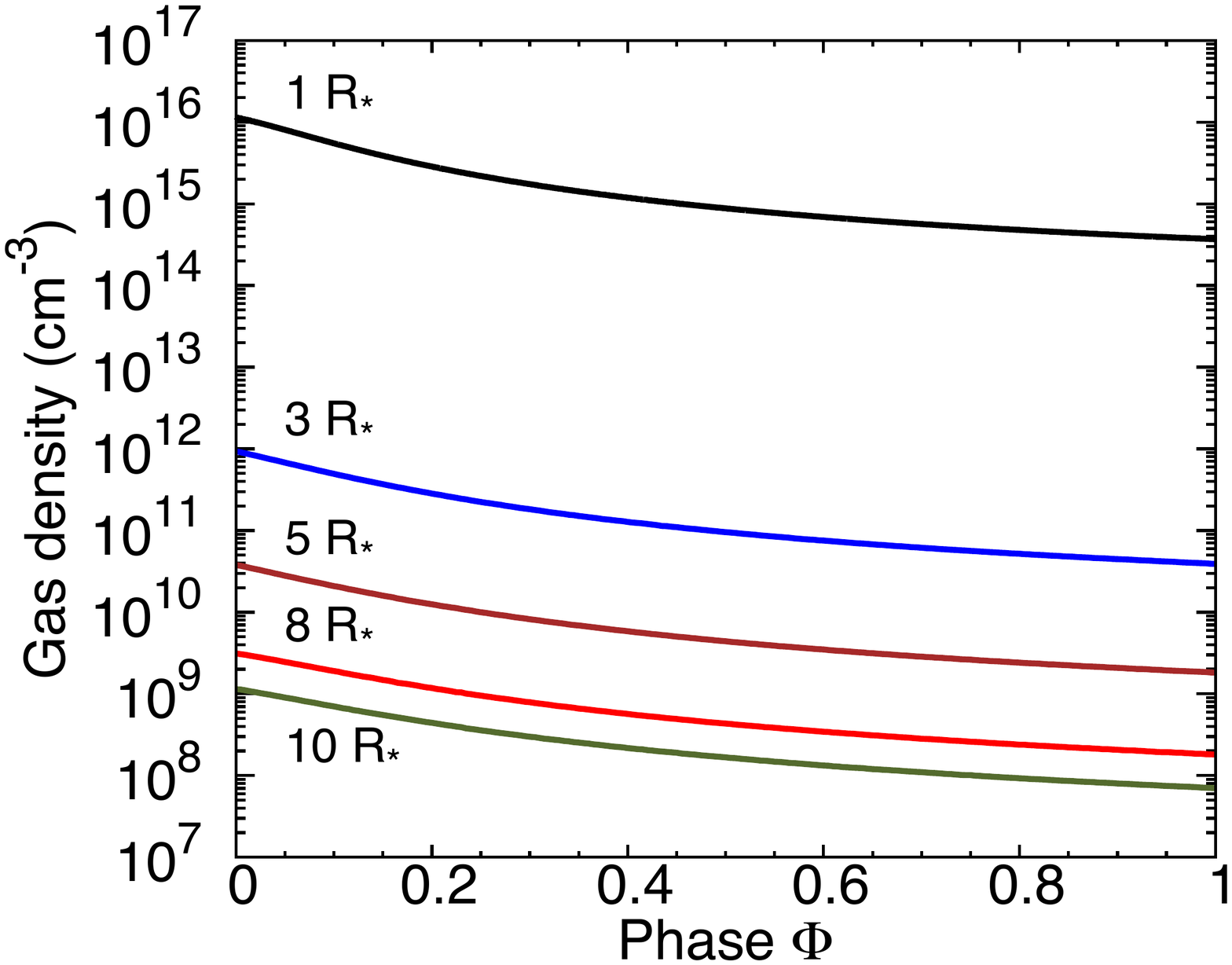} 
   
   \caption{Gas temperature and number density variation in the post-shock gas over one pulsation period as a function of phase for radius between 1 and 10~\rs.}
 \label{fig2}
\end{figure}

For IK Tau, several shock velocities were tested on the gas phase chemistry (20, 25 and 32 \kms), and the best results were achieved for the stronger shock of 32~\kms at $r_s=1$~\rs, as in the IK Tau study of Duari et al. (1999). Owing to the conservation of energy, the shock strength $V_s$ decreases as the shock crosses the inner wind according to energy conservation and $V_s (r)~=~V_s(r_s)\sqrt{r_s/r}$. The stellar and inner wind parameters are summarised in Table \ref{tab1}. 
We first consider a gas layer at $r_1$ that is shocked, moves forwards, and falls back to its initial position owing to the stellar gravity. We run the chemistry and grain condensation over two full pulsation periods to check for periodicity and then push this gas layer to a larger radius $r _2= r_1 + \Delta r$ characterised by a lower pre-shock gas density and temperature. The outputs at $r_1$, specifically the molecular abundances, as well as the abundance of clusters and dust grains, are used as inputs to the calculations at radius $r_2$. We run the model of the gas phase, the nucleation of dust clusters and their condensation in dust grains from $r_s = 1$~\rs\ to 10~\rs. The gas parameters from 1~\rs\ to 10~\rs\ are listed in Table \ref{tab2}, while the variation of gas temperature and density over a pulsation phase and between 1~\rs\ and 10~\rs\ are illustrated in Figure \ref{fig2}. 
We direct the reader to the papers by Cherchneff et al. (1992), Cherchneff (1996), Willacy \& Cherchneff (1998), Duari et al. (1999), Cau (2002) and Cherchneff (2006, 2011, 2012) for full details of the formalism and model we use to describe the inner wind of AGB stars. 
 
This simple analytical model does not represent a fully consistent picture of the shocked upper stellar atmosphere and does not aim to reproduce mass-loss rates in AGB stars. We do not attempt to estimate the radiation pressure force acting on newly formed dust grains in the gas and derive a mass loss for IK Tau. We consider strictly ballistic motions of the gas above the photosphere and identify the location of cluster formation and dust condensation in the inner wind. Therefore, our model provides a simplified description compared to more sophisticated hydrodynamic models of the inner wind, which take into account the dust radiative feedback on the gas (e.g., \cite{win00, hof03, woit06}). While such models aim to reproduce the mass loss rates and optical properties of AGB winds, they do not seem to explain the non-equilibrium chemical processes that trigger the formation of molecules, dust clusters and the growth of dust grains from the gas phase. Indeed, they often assume over-simplifying mechanisms (e.g., thermodynamic equilibrium) for the production of molecules and dust grains. Our simplified physical model of the shocked layers above the stellar photosphere reproduces to some extent the shocked gas conditions, as shown by Bertschinger \& Chevalier (1985), and is thus sufficient to focus on and understand the chemical processes at play in the dust formation zone.

\begin{table*}
\caption{Chemical species and dust clusters included in the IK Tau chemical model..}              
\label{tab3}      
\centering                         
\begin{tabular}{ll  ll ll  ll  ll}         
\hline\hline  
 Atoms & \multicolumn{7}{c}{Molecules} & &\\
 \hline               
O & OH & H$_2$O & O$_2$ & &  & & & & \\ 
C & CH & CO & CO$_2$ &CS & OCS & HCN  & CN  & & \\
Si & SiO &  HSiO &  H$_2$SiO & SiO$_2$ & SiS & Si$_2$  &\\
N & NH &  NO & NH$_2$& NH$_3$ &N$_2$ & & \\
S & SH & H$_2$S & SO  & SO$_2$ & HOSO &  & & & \\
P & PH &P$_2$ & PO& PN&  & & &\\
Mg & MgH &MgO & MgOH&  & & & &\\
Fe &FeH  & FeO&  Fe$_2$& & & & &\\ 
F& HF & F$_2$   & &  & & & &\\ 
Al & AlH & AlO &AlOH & AlO$_2$& AlCl & Al$_2$  & Al$_2$O& &\\ 
Cl & HCl &Cl$_2$ & ClO &  && & &\\
Ca & CaO & CaOH& CaCl & & & & & & \\
Na &  NaH& NaCl &   & & & & &\\ 
Ti & TiO & TiO$_2$ & & & & & & & \\
K & KH& KCl &   & & & & &\\ 
H &H$_2$ & & & & & & & & \\
\hline
\multicolumn{10}{c}{{Dust clusters}} \\
\hline
Si$_2$O$_2$ & Si$_2$O$_3$& Si$_3$O$_3$ & Si$_4$O$_4$ & H$_2$Si$_2$O$_2$ & HSi$_2$O$_3$ & H$_2$Si$_2$O$_3$ & H$_2$Si$_2$O$_4$ & & \\
MgSi$_2$O$_3$ & MgSi$_2$O$_4$ & MgSi$_2$O$_5$ & MgSi$_2$O$_6$H$_2$ &Mg$_2$Si$_2$O$_4$ & Mg$_2$Si$_2$O$_5$ & Mg$_2$Si$_2$O$_6$ & Mg$_2$Si$_2$O$_7$H$_2$ & & \\
Mg$_3$Si$_2$O$_5$ & Mg$_3$Si$_2$O$_6$ & Mg$_3$Si$_2$O$_7$ & Mg$_3$Si$_2$O$_8$H$_2$ & Mg$_4$Si$_2$O$_6$ & Mg$_4$Si$_2$O$_7$ & Mg$_4$Si$_2$O$_8$  & & &  \\
Al$_2$O$_2$ &  Al$_2$O$_3$ & Al$_4$O$_6$ & & & & & & & \\
\hline  
\end{tabular}
\end{table*}

\subsection{Chemical kinetic model and condensation of dust grains}
\label{sec3}

Photospheric atomic and molecular abundances are calculated by assuming TE at 1~\rs\ and by taking the initial elemental composition given by the FRUITY database of Cristallo et al. (2011), for a 1.5~\Ms\ star on the ZAMS with solar metallicity, that has experienced three thermal pulses on the AGB. The derived stellar mass M$_{\star}$ and C/O ratio after these three thermal pulses are 1~\Ms\ and  0.6, respectively. The C/O value is close to the assumed C/O ratio of 0.75 used in Duari et al. (1999) and Cherchneff (2006). The effect of using this lower C/O ratio value as opposed to 0.75 has been assessed and is minor on all molecular and cluster abundances. We then choose to keep the C/O ratio of 0.75 in order to better compare our results with previous models of IK Tau.

\subsubsection{Chemistry} 

The rich chemistry of the IK Tau inner wind is modelled by a network of chemical reactions that includes 100 species. The species were chosen according to the following criteria: a) astronomical detection in AGB and supergiant winds, b) association with the chemistry of detected molecules, and c) implication in the nucleation of dust clusters. The list of species is given in Table \ref{tab3}.  

All possible types of reactions, effective in a high temperature and density gas, are considered. They include termolecular processes that are active at high gas densities, and bimolecular processes, which include neutral exchange and radiative association reactions, and collisional dissociation. The chemical networks of Duari et al. (1999) and Cherchneff (2006) have been revised and extended to include the processes involved in the formation of clusters (see below and Sarangi \& Cherchneff 2013, 2015). Reaction rates were taken primarily from the National Institute of Standards and Technology Chemical Kinetics Database, the 2012 version of the UMIST Database for Astrochemistry (\cite{mcel13}), specific studies of ceramic production in flames (e.g., \cite{zac95}) and modelling of other circumstellar environments (Cherchneff \& Dwek 2009, 2010, \cite{sar13}). 

Following Cherchneff (2011a, 2012), the implementation of rates for reverse reactions was changed to accommodate available new rates measured in combustion and aerosol chemistry. The rate of a specific reverse process was not calculated from the equilibrium constant by assuming detailed balance as in Duari et al. (1999) and Cherchneff (2006), but directly entered in the chemical network with its available measured, calculated, or estimated value. This new, purely kinetic approach has been tested for the case of the carbon star IRC+10216 and provides better results when compared to observational data. Specifically, the formation of water and other relevant molecules in carbon stars was successfully explained by this revised shock-induced chemistry applied to the inner stellar wind (\cite{cher11a, cher12}). 

No ionic species and related chemical reactions are considered because we assume there are no ionisation processes of atomic and molecular species induced by the ultraviolet stellar radiation field, which is low for the effective temperature of IK Tau. Furthermore, we assume there is no penetration of the interstellar radiation field or cosmic rays in these regions close to the stellar surface. 

\subsubsection{Dust clusters}

\begin{table}
\caption{Chemical reactions that initiate the formation of silaformyl and the silanone dimerisation, adapted from Zachariah \& Tsang (1995), and the prevalent reactions leading to the formation of enstatite dimer, adapted from Goumans \& Bromley (2012). }           
\label{tab4}       
\centering                         
\begin{tabular}{l l l l l l l l}        
\hline\hline      
R1 & SiO& +& H & $ \leftrightarrow$ & HSiO & & \\   
R2&SiO& +& H$_2$ & $ \leftrightarrow$ & HSiO & +&H\\   
R3&HSiO &+ &H& $ \leftrightarrow$ &H$_2$SiO & & \\
R4&HSiO &+ &H$_2$& $ \leftrightarrow$ &H$_2$SiO &+& H\\
R5&H$_2$SiO&+ &H$_2$SiO& $ \leftrightarrow$ &H$_2$Si$_2$O$_2$ &+ &H$_2$\\
R6&H$_2$Si$_2$O$_2$ &+ &H$_2$O &$ \leftrightarrow$ &H$_2$Si$_2$O$_3$ &+& H$_2$\\
R7&H$_2$Si$_2$O$_3$ &+& Mg & $ \leftrightarrow$ & MgSi$_2$O$_3$ &+& H$_2$ \\
R8&MgSi$_2$O$_3$ &+ &H$_2$O &$ \leftrightarrow$ &MgSi$_2$O$_4$ &+& H$_2$\\
R9&MgSi$_2$O$_4$ & + & Mg &$ \leftrightarrow$ &Mg$_2$Si$_2$O$_4$ & & \\
R10&Mg$_2$Si$_2$O$_4$ &+ &H$_2$O &$ \leftrightarrow$ &Mg$_2$Si$_2$O$_5$ &+& H$_2$\\
R11&Mg$_2$Si$_2$O$_5$ &+ &H$_2$O &$ \leftrightarrow$ &Mg$_2$Si$_2$O$_6$ &+& H$_2$\\
\hline                                    
\end{tabular}
\end{table}

Together with the formation of molecules, we model the synthesis of molecular clusters that are precursors to dust grains. We consider two types of prevalent condensates in the inner wind, silicates and alumina, and consider as dust clusters the dimers of forsterite (\fors) and enstatite (\ens), and the dimer of \al. The formation routes of silicate dimers of forsterite and enstatite stoichometry were studied by Goumans \& Bromley (2012). By identifying the structures of particularly energetically stable small clusters, their thermodynamic properties were estimated and used to highlight potential cluster formation routes at equilibrium. For the gas parameters illustrated in Figure \ref{fig2}, the exergonocity of reactions involving clusters is assessed by using the thermodynamic data estimated by Goumans \& Bromley (2012) for the clusters listed in Table \ref{tab3}, and reaction rates for both formation and destruction processes are estimated on the basis of known rates for similar reaction types and the likelihood of the reactions to proceed. The synthesis of silicate clusters is described by the recurrent addition of oxygen to the cluster lattice through reaction with H$_2$O, and the inclusion of one Mg atom after the oxidation step. 

This suite of reactions leads to the formation of enstatite (\ens) and forsterite (\fors) dimers. In the study by Goumans \& Bromley, the growth to silicate small clusters follows a first important step, the dimerisation of SiO, which acts as a bottleneck to the growth mechanism. However, the dimerisation of SiO is a slow process. No measured rate value exists, but the theoretical study by Zachariah \& Tsang (1993) provides estimates of SiO addition rates up to the SiO pentamer, and their reverse dissociative paths, and these rates are used in this study. Because of the overwhelming abundance of H and H$_2$ in the inner wind, another pathway to the formation of small Mg/Si/O clusters is included that involves the dimerisation of the silaformyl radical, HSiO, followed by the formation of silanone, H$_2$SiO, and its dimerisation. These chemical routes were studied by Zachariah \& Tsang (1995) in the context of the nucleation of silica in a silane-rich flame. We list in Table \ref{tab4} the chemical reactions involved in the nucleation of the MgSi$_2$O$_3$ molecular cluster (R1$-$R7). The later growth of small silicate clusters resumes with the mechanism derived by Goumans \& Bromley, and the oxygen addition to MgSi$_2$O$_3$ through reaction with the prevalent O-bearing species, H$_2$O, according to Reactions R8$-$R11.

The formation of alumina dimers, (Al$_2$O$_3$)$_2$, is described as follows: the formation of \al\ results from the trimolecular dimerisation of AlO, followed by the oxidation of (AlO)$_2$ through reaction with O-bearing species, essentially H$_2$O. This choice was based on similarities in the structure of Si$_2$O$_3$ and Al$_2$O$_3$ as both molecules have kite-shaped, global minimum structures (\cite{pat99,li12}). Because no measurements appear to exist on the  gas-phase formation routes to Al$_2$O$_3$, we have assumed chemical routes and related rates similar to those for the formation of Si$_2$O$_3$ and H$_2$Si$_2$O$_3$. 

\subsubsection{Condensation of dust grains} 
Simultaneously to the formation of molecules and dust clusters from the gas phase by non-equilibrium chemistry, we model the condensation of dust grains following Plane (2013) and Sarangi \& Cherchneff (2015). We choose the forsterite and alumina dimers, (\fors)$_2$ and (\al)$_2$, respectively, as stable condensation nuclei, and consider both the coalescence and coagulation of these dimers to form dust grains. The formalism is based on a Brownian formalism, which accounts for the collision, scattering, coalescence and coagulation of the grains through Brownian motion. 
We use the equation formulated by Jacobson (2005), whereby the variation with time of the number density of a grain of specific volume $v$ is described by the integro-differential coagulation equation given by 

\begin{equation} 
\label{eq1}
\frac{dn_v (t)}{dt} = \frac{1}{2} \int\limits_0^v \beta_{v-v',v} n_{v-v'}n_{v'} dv' - n_{v} \int\limits_0^\infty \beta_{v,v'} n_{v'} dv'.
\end{equation}
Here, $t$ is the time, $v'$ and $(v-v')$ are the volumes of the two coagulating particles, n$_v$ is the number density of grains with volume $v$, and  $\beta_{v, v'}$ is the rate coefficient of coagulation between particles with volume $v$ and $v'$. The shocked gas is characterised by a free-molecular regime, defined by the grain mean-free path in the gas being much larger than the grain size, where Brownian diffusion prevails in the coagulation process and controls the rate of coagulation $\beta_{v_i,v_j}$ for particles $i$ and $j$. Equation \ref{eq1} was solved by using a semi-implicit volume conserved model, where grains are assumed spherical (\cite{sp06}). For both alumina and silicate, grains are assigned to discrete bins, following a volume ratio distribution given by $v_n = \gamma^{n-1} v_0$, where $v_n$ is the volume of the $n^{th}$ bin, $\gamma$ is a constant defined as the ratio of the volumes of adjacent bins, and $v_0$ is the volume of the first bin, determined by the size of the gas-phase precursor. At each time step, the volume of the first bin, $v_0$ corresponds to the radius $a_0$ of the stable, largest clusters produced from chemical kinetics in the nucleation phase (for alumina and silicate, $a_0=3.45$~\AA~and $3.3$~\AA, respectively - see \cite{sar15}). The grain size distributions are then derived for silicates of forsterite and enstatite stoichiometry, and alumina. 
We direct the reader to Plane (2013) (and references therein) and Sarangi \& Cherchneff (2015) for a detailed description of the condensation formalism, the expressions for the rates of Brownian diffusion and coalescence, and their variation with gas parameters. 

Our formalism for grain growth does not consider deposition of atoms and molecules on the surface of small grains. Once the radius of grains reaches several tens of \AA, the growth of the particles proceeds through surface deposition, which is a fast process, and coagulation and coalescence, as observed in flames (\cite{ro07}). For deposition, the sticking coefficient $S$ of an atom or molecule on the surface of grains depends on the chemical composition of the accreting species and the grain surface, and the surface and gas temperature. The quantity may vary over orders of magnitude depending on the grain material, with value range of $\sim 10^{-4}- 1$ (e.g., \cite{lei85, bru93, nag96}). Existing studies on dust formation in circumstellar environments often assume a growth process through surface deposition only, and an $S$ value equal to unity or in the range $0.1-1$ (e.g., \cite{gail99, bla15}). In view of these uncertainties, we prefer to not consider surface deposition in the formalism at this stage of our study, and discuss the potential effects of accretion on the present results in \S~\ref{dust}. 

\section{Results}
\label{sec4}

The derivation of molecular and cluster abundances is as follows: we assume that the initial composition of the pre-shocked photosphere is given by TE calculations for the parameters in Table \ref{tab1}. We derive the abundances of molecules and clusters in the immediate ``fast chemistry'' post-shock gas and the``adiabatic trajectory'' of a gas parcel (see \cite{wil98}) by solving a system of 100 master equations described by a chemical network that simultaneously includes the formation of chemical species, and alumina and silicate clusters. When the clusters have reached the size of dimers of alumina and forsterite, they start to grow by coalescence and coagulation. As explained in \S \ref{physmod}, the parcel of gas is then lifted from radius $r$ to radius $r'$, and the calculated abundances for the shocked gas at $r_1$ are then used as initial pre-shock conditions for the gas parcel at $r_2=r_1 + \Delta r$, with $\Delta r = 0.5$~\rs. The radius $r_1$ was then varied from 1~\rs\ to 10~\rs. We first discuss the results of the gas phase in \S~\ref{gas} and then present results on dust clusters and dust grains in \S~\ref{dust}. 

\subsection{Gas phase}
\label{gas}
\begin{figure}
\centering
  \includegraphics[width=\columnwidth]{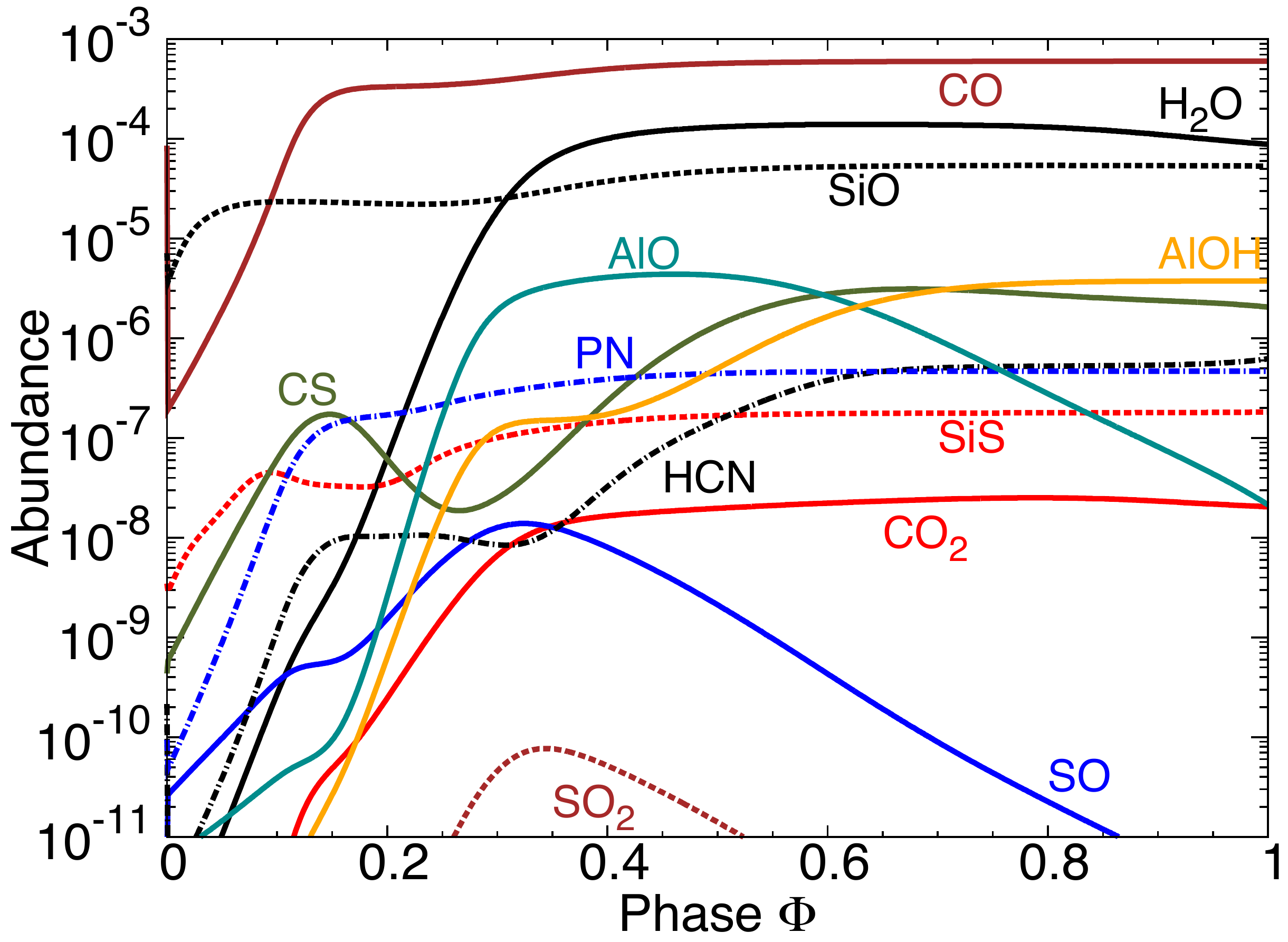}
  \includegraphics[width=\columnwidth]{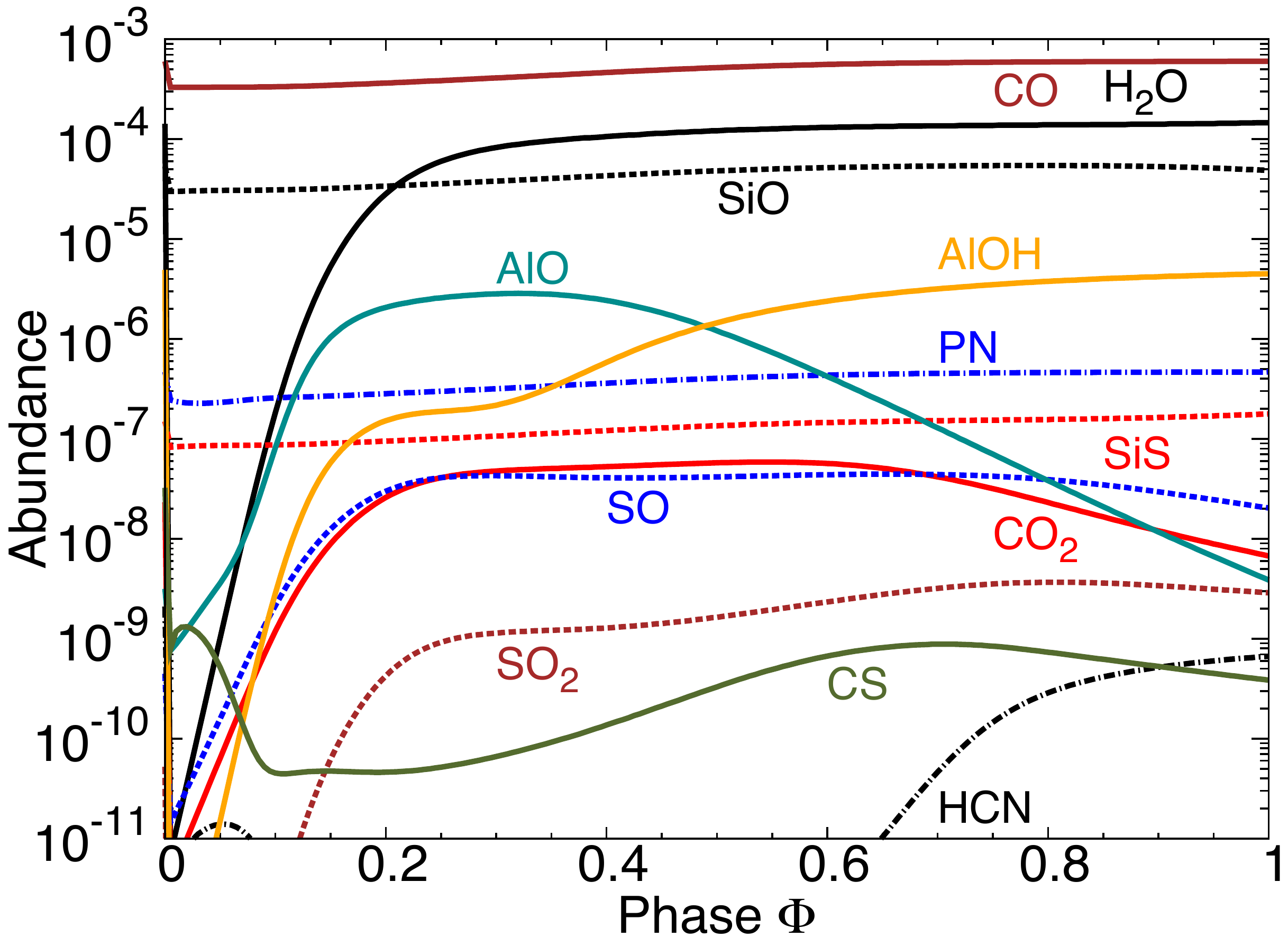}
\caption{Molecular abundances with respect to total gas number density n$_{gas}$ as a function of pulsation phase for two specific radii in the IK Tau inner wind. Top: 1~\rs; Bottom: 3~\rs.}
\label{fig3}
\end{figure}

\begin{figure}
\centering
  \includegraphics[width=9.22cm, height=7.2cm]{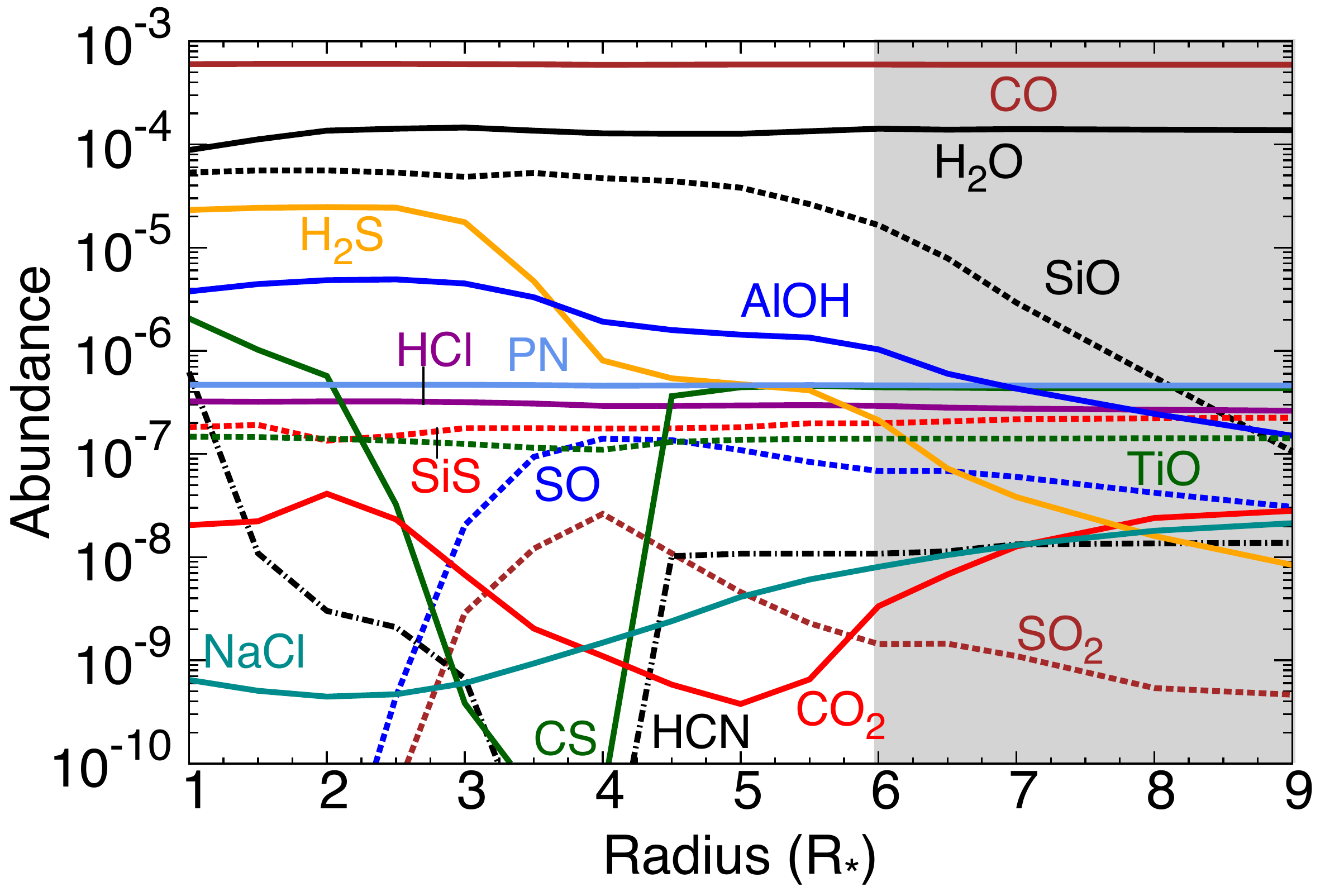}
\caption{Molecular abundances with respect to total gas number density n$_{gas}$ as a function of radius in the inner wind of IK Tau. Values correspond to $\phi=1$ in the gas layer and the grey shaded area corresponds to the region where the wind is already accelerated - see \S~\ref{sili}.   }
\label{fig4}
\end{figure}

Molecular abundances versus pulsation phase at two radii in the inner wind are shown in Figure \ref{fig3}. At 1~\rs, equivalent to the shocked stellar photosphere position, the shock destroys all molecules in the H$_2$ dissociation zone, and species gradually reform in the cooling region until  $\phi \simeq 0.2-0.4$, depending on molecules. Beyond this phase, some molecules are fully reformed with almost constant abundances, such as CO, SiS, and PN, while others are still chemically active, as for example, SO, and AlO. At larger radius, the early destruction of molecules in the H$_2$ dissociation zone still occurs, but it is less efficient. Some molecules maintain almost similar abundances to those at 1~\rs, e.g., SiS, SiO, and PN, while others are still chemically active (e.g., AlO). The prevalent molecules formed in the inner wind are illustrated in Figure \ref{fig4}, where abundances with respect to total gas as a function of radius are shown, at a phase $\phi=1$. As previously mentioned, some molecules form in the post-shock gas of the shocked photosphere and maintain abundances almost constant in the inner wind. These species include CO, PN, HCl, TiO, and SiS. The rather constant abundances reflect the fact that those species do not participate in complex chemistries and in the formation of dust clusters in the post-shock gas. Once they form in the shocked gas at 1~\rs, they stay unchanged until the chemistry is frozen out at the wind acceleration radius after $r > 6$~\rs\ (see \S~\ref{sili}). 

Conversely, other molecules actively partake in the chemistry of the gas phase, including that of dust cluster formation. This is the case for H$_2$S, SO, AlOH, SiO, and the carbon-bearing species CO$_2$, HCN and CS. While SiO and AlOH, through AlO, are directly associated with the nucleation of silicates and alumina, respectively, other molecules are linked by the shock-sensitive chemistry, like H$_2$S and SO. The formation of H$_2$S occurs close to the star through the reaction 

\begin{equation}
\label{eq1}
{\rm SH + H_2 \rightarrow H_2S + H,}
\end{equation}
where SH forms in the hot post-shock gas through the reaction 
\begin{equation}
\label{eq2}
{\rm S + H_2 \rightarrow SH + H,}
\end{equation}
which has a high activation energy. At larger radii and for weaker shocks and lower post-shock temperatures, the formation of SO proceeds essentially through
\begin{equation}
\label{eq3}
{\rm S + OH \rightarrow SO + H,}
\end{equation}
for which the only rate available is at 300K, and 
\begin{equation}
\label{eq4}
{\rm SH + O \rightarrow SO + H.}
\end{equation}
The rates for reactions \ref{eq3} and \ref{eq4} have no activation energy barrier and can easily proceed in the inner wind to form SO. However, the reverse reactions have a high activation energy ($\sim 11000$ K and $20000$ K for the reverse of reactions \ref{eq3} and \ref{eq4}, respectively), and thus proceed close to the star and inhibit the formation of SO at small radii. Through reaction \ref{eq4}, SO formation depletes the SH reservoir and then hampers the formation of H$_2$S, whose abundance starts dropping around 3~\rs. The production of SO triggers the formation of SO$_2$ from the gas phase following the reaction 
\begin{equation}
\label{eq5}
{\rm SO + OH \rightarrow SO_2 + H,}
\end{equation}
which is barrierless and can proceed as soon as the SO abundance builds up. However, we see from Figure \ref{fig4} that the SO$_2$ abundance remains always low in the inner wind, and much lower than that of SO by almost two orders of magnitude. 

\begin{figure}
\centering
  \includegraphics[width=\columnwidth]{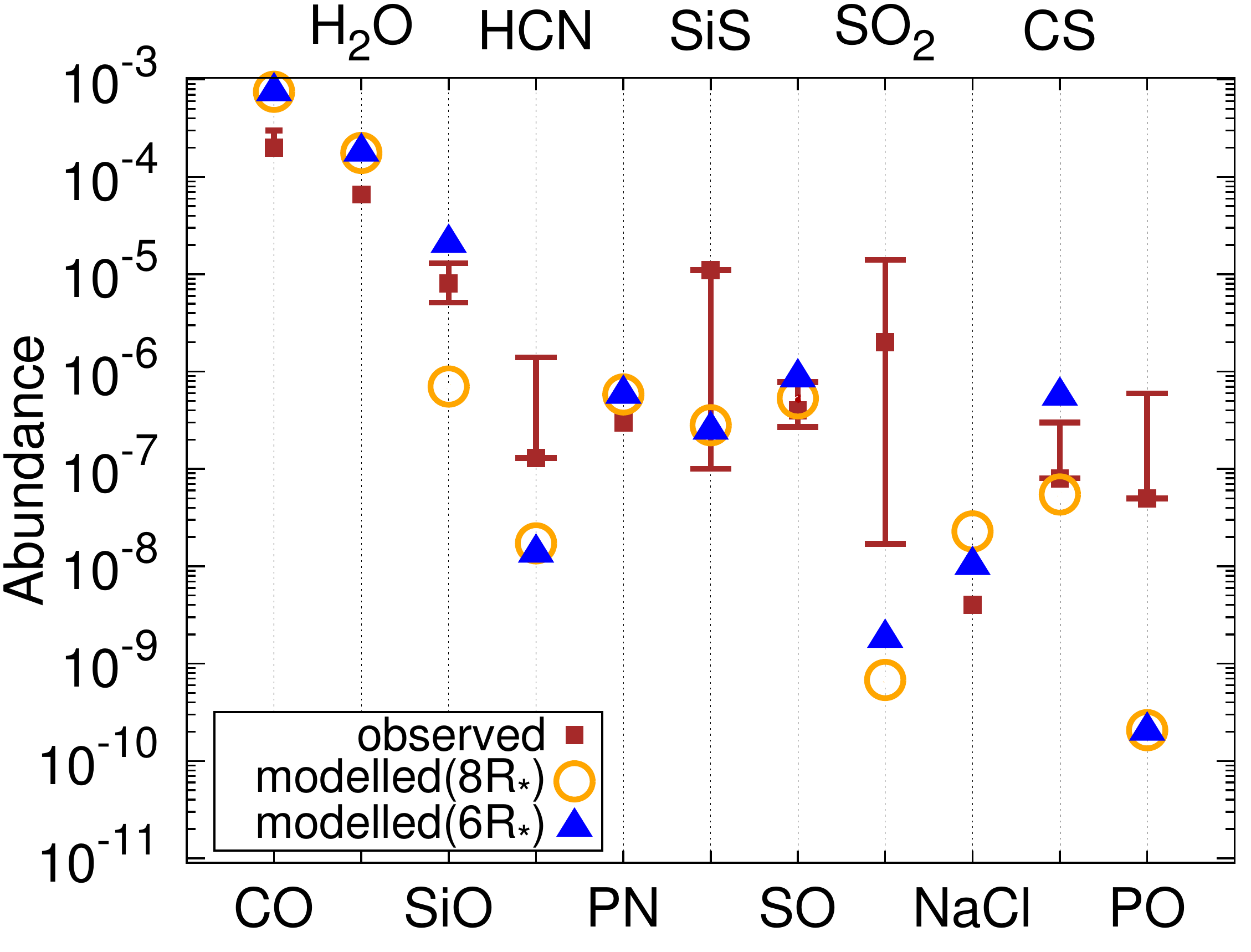}
\caption{Modelled and observed molecular abundances in the inner wind of IK Tau. The modelled values are for $\phi=1$, and $r = 6$~\rs\ and $8$~\rs. The bars for observed data indicate the value range derived from the different sets of observations listed in Table \ref{tab5}.}
\label{fig5}
\end{figure}

The C-bearing species, HCN and CS, form from a complex chemistry that is linked to the SH/H$_2$S chemistry. Basically, the dominant formation processes for CS are
\begin{equation}
\label{eq6}
{\rm C + SH \rightarrow CS + H,}
\end{equation}
\begin{equation}
\label{eq7}
{\rm S + CN \rightarrow CS + H,}
\end{equation}
and 
\begin{equation}
\label{eq8}
{\rm CO + SH \rightarrow OCS + H,}
\end{equation}
followed by 
\begin{equation}
\label{eq9}
{\rm OCS + H \rightarrow CS + OH. }
\end{equation}
Therefore, the CS, CN, SH and H$_2$S molecules are chemically linked. Inspection of Figure \ref{fig4} indicates a dip in the CS and HCN abundances between $2$~\rs\ and $4 $~\rs. In that radius range, the formation of silicate dust clusters that is initiated by the formation of HSiO and H$_2$SiO depletes molecular hydrogen to release atomic hydrogen in the process. This excess favours the reverse of reaction \ref{eq2}, producing copious amounts of sulphur atoms and decreasing the SH and H$_2$S reservoirs. Therefore, the CS abundance drops as reactions \ref{eq6}, \ref{eq8} and \ref{eq9} become less efficient. The CN radical initially forms in the H$_2$ dissociation region after the shock front from the reactions
\begin{equation}
\label{eq10}
{\rm N + CO \rightarrow CN + O,} 
\end{equation}
and 
\begin{equation}
\label{eq11}
{\rm N + CS \rightarrow CN + S. }
\end{equation}

The formation of HCN occurs in all post-shock regions of the inner wind following the reaction
\begin{equation}
\label{eq11a}
{\rm CN + H_2 \rightarrow HCN + H}
\end{equation}

Both CN and HCN abundances are regulated by reaction \ref{eq11a} and its reverse process, and the formation of HCN is then controlled by the H$_2$ molecule in the inner wind. Therefore, the drop in the HCN abundance ensues from the depletion of H$_2$ due to the nucleation of silicates between $2$~\rs\ and $4$~\rs. 

Finally, the molecule CO$_2$ was detected with the Infrared Space Observatory, ISO, in two Mira stars, T Cep (\cite{yam99}) and R Cas (\cite{mark00}), but has not been detected in IK Tau to date. We predict the formation of CO$_2$ by the reaction 
\begin{equation}
\label{eq11}
{\rm CO + OH \rightarrow CO_2 + H, }
\end{equation}
which is active in O-rich AGB winds (\cite{cher06}). This reaction goes backwards when the synthesis of silicate clusters occurs and replenishes the gas in atomic H at the expense of H$_2$ in the $3-6$~\rs\ region. A slight drop in CO$_2$ abundance then occurs, as shown in Figure \ref{fig4}. 

The modelled molecular abundances for IK Tau are summarised in Table \ref{tab5}, where abundances in the stellar photosphere at TE are listed, and the abundances in the shocked wind are given at 1~\rs\ and 6~\rs. These values are compared to abundances derived from the modelling of observational data of the IK Tau inner wind. We see that two distinct groups of molecules exist. The first group includes species that are present at TE in the photosphere, are destroyed by shocks, but reform with almost their TE abundance values (e.g., H$_2$O, SiO, SiS, SO, CO$_2$, TiO, and HCl). The second group includes species that are not present in the photosphere and only form in the inner wind as a result of shock chemistry (e.g., HCN, CS, PN, AlOH, NaCl). As for the agreement between modelled and observed abundance values, we consider that both sets of values agree well when the discrepancy is within a factor of 10. In Figure \ref{fig5}, we plot the modelled abundances at 6~\rs\ and 8~\rs\ and the values derived from observations. We see that both modelled and observed sets of data agree well for 9 species out of 11 molecules detected. On the other hand, the modelled abundance values for SO$_2$ and PO fail to agree with observations. 

\begin{table*}
\caption{Predicted abundances (with respect to H$_2$) of selected molecules. TE abundances are given at r$_s$ and modelled abundances are listed for r$_s$ and 6~\rs. Most recent abundances (with respect to H$_2$) of molecules observed in the inner wind of IK Tau and other objects are listed, with the corresponding references. The double horizontal line separates data available for IK tau (upper part) from data on other objects (lower part). }             
\label{tab5}     
\centering                          
\begin{tabular}{l l l l c l }       
\hline\hline                

Species & TE at $r_s$& Modelled at r$_s$& Modelled at $6$~\rs& Observed & Reference \\
 \hline
CO     & $1.0 \times 10^{-3}$  & $7.3 \times 10^{-4}$  & $7.4 \times 10^{-4}$  & $2.0 \times 10^{-4}$ & \cite{dec10b} \\
 &  && & $ 3.0 \times 10^{-4}$ & \cite{kim10} \\
 \hline
H$_2$O & $2.1 \times 10^{-4}$  & $1.1 \times 10^{-4}$  & $1.8 \times 10^{-4}$ & $6.6 \times 10^{-5}$ & \cite{dec10b}\\
\hline
SiO    & $9.5 \times 10^{-5}$  & $6.4 \times 10^{-5}$  & $1.9 \times 10^{-5}$ & $8.0 \times 10^{-6}$ &\cite{dec10b} \\
 & -- &-- & -- & $ 5.1 \times 10^{-6} - 1.3 \times 10^{-5}$ &\cite{kim10} \\
\hline
SiS    & $4.0 \times 10^{-8}$  & $2.2 \times 10^{-7}$  & $2.7 \times 10^{-7}$ & $1.1 \times 10^{-5}$ & \cite{dec10a} \\
   & -- &-- & -- & $3.1 \times 10^{-7} - 1.3 \times 10^{-6}$ & \cite{kim10} \\  
   &-- &--& -- & $1.0 \times 10^{-7}$& \cite{scho07} \\  
   \hline
SO     & $4.3 \times 10^{-8}$  & $2.1 \times 10^{-12}$ & $7.3 \times 10^{-8}$ & $4.0 \times 10^{-7}$  & \cite{dec10a} \\
&- - &-- & -- & $2.7 \times 10^{-7} - 7.8 \times 10^{-7}$ & \cite{kim10} \\ 
\hline

SO$_2$ & $1.3 \times 10^{-11}$ & $1.8 \times 10^{-14}$ & $4.3 \times 10^{-9}$ & $2.0 \times 10^{-6}$& \cite{dec10a}\\
& -- &-- & -- & $4.2 \times 10^{-6} - 1.4 \times 10^{-5}$  & \cite{kim10} \\ 
  & -- &-- & -- & $1.7 \times 10^{-8} - 3.7 \times 10^{-7}$  \tablefootmark{a} & \cite{yam99} \\  
\hline
HCN   & $6.3 \times 10^{-12}$ & $7.7 \times 10^{-7}$  & $5.0 \times 10^{-8}$ & $1.3 \times 10^{-7}$ & \cite{scho13} \\  
& -- &-- & -- & $4.3 \times 10^{-7} - 1.4 \times 10^{-6}$ & \cite{kim10} \\  
& -- &-- & -- & $4.4\times 10^{-7} $ & \cite{dec10b} \\
\hline
CS  & $1.9 \times 10^{-11}$ & $2.5 \times 10^{-6}$  & $5.7 \times 10^{-7}$   & $8.0 \times 10^{-8}$  & \cite{dec10b} \\
& -- &-- & -- & $8.1 \times 10^{-8} - 3.0 \times 10^{-7}$ & \cite{kim10} \\  
\hline
PN     & $4.0 \times 10^{-10}$ & $5.7 \times 10^{-7}$  & $5.8 \times 10^{-7}$  & $3.0 \times 10^{-7}$ & \cite{deb13} \\
\hline
PO   & $1.0 \times 10^{-7}$  & $3.3 \times 10^{-10}$ & $2.0 \times 10^{-10}$ & $5.4\times 10^{-8} - 6.0 \times 10^{-7}$ & \cite{deb13} \\
\hline
NaCl   & $3.7 \times 10^{-12}$ & $7.8 \times 10^{-10}$ & $1.0 \times 10^{-8}$  & $4.0 \times 10^{-9}$ & \cite{mil07} \\
\hline
NH$_3$ &$1.1 \times 10^{-11}$&$2.8\times 10^{-9}$&$1.5\times 10^{-19}$ & $1-3 \times 10^{-6}$& \cite{men10} \\
\hline
\hline
CO$_2$ & $4.1 \times 10^{-8}$  & $2.5 \times 10^{-8}$  & $4.4 \times 10^{-9}$ & $5.9 \times 10^{-8}$ \tablefootmark{b} &
 \cite{mark00}  \\
  & -- &-- &-- & $2 \times 10^{-9} - 4 \times 10^{-8}$  \tablefootmark{a} & \cite{yam99} \\ 
AlO & $8.0 \times 10^{-9} $& $ 2.6 \times 10^{-8} $& $1.6 \times 10^{-10} $ & $3.4 \times 10^{-8} $\tablefootmark{c}  & \cite{kam13a} \\
AlOH   & $4.5 \times 10^{-9}$  & $4.5 \times 10^{-6}$  & $1.5 \times 10^{-6}$  & $1.0 \times 10^{-7}$ \tablefootmark{c} & \cite{ten10} \\
TiO \tablefootmark{d} & $2.4 \times 10^{-7}$  & $1.8 \times 10^{-7}$  & $1.8 \times 10^{-7}$  & $ 5.0 \times 10^{-9} $  \tablefootmark{c} & \cite{kam13b} \\
TiO$_2$ & $1.3 \times 10^{-12}$  & $6.2 \times 10^{-10}$  & $4.8 \times 10^{-9}$  & $6.3 \times 10^{-10}$ \tablefootmark{c}& \cite{kam13b} \\
 & -- &-- &-- &  $3.8 \times 10^{-8}$ \tablefootmark{c}& \cite{deb15} \\
HCl & $3.5 \times 10^{-7}$  & $3.9 \times 10^{-7}$  & $3.7 \times 10^{-7}$  & $1.5 \times 10^{-8}$ \tablefootmark{e}& \cite{yam00} \\
AlCl  & $1.5 \times 10^{-10}$  & $3.8 \times 10^{-12}$  & $2.2 \times 10^{-10}$  & $ < 5 \times 10^{-8}$ \tablefootmark{c}& \cite{ten10} \\
SH     & $2.8 \times 10^{-6}$  & $1.3 \times 10^{-7}$  & $2.3 \times 10^{-8}$  & $2.0 \times 10^{-7}$ \tablefootmark{e} & \cite{yam00}\\
H$_2$S & $8.1 \times 10^{-8}$  & $2.8 \times 10^{-5}$  & $1.0 \times 10^{-8}$  & $ < 2\times 10^{-6}$ \tablefootmark{e}  & \cite{yam00}  \\
\hline
\end{tabular}
\tablefoot{
\tablefoottext{a}{SO$_2$ \& CO$_2$ abundances derived from data of T Cephei}
\tablefoottext{b}{The [CO$_2$]/[H$_2$O] ratio as measured in R Cas was used to estimate CO$_2$ abundance in IK Tau}
\tablefoottext{c}{Measured in the supergiant VY Canis Majoris}
\tablefoottext{d}{Also detected in the inner wind of the S star NP Aurigae by Smolders et al. (2012)}
\tablefoottext{e}{Measured at 1.1~\rs\ in the S star R Andromedae.}
}
\end{table*}

For SO$_2$, we find a low abundance of $\sim 2 \times 10^{-9}$ at 6~\rs, which indicates that the molecule does not efficiently form in the gas phase from the reaction of SO with OH. Yamamura et al. (1999) detected the $7.3$~\mic\ band of SO$_2$ in the ISO/SWS spectra of the Mira star T Cep, and derived an SO$_2$ abundance in the range $2\times 10^{-8} - 4\times 10^{-7}$ in a shell located at $4-5$~\rs. Most of the IR band emission comes from the inner layers, and we notice that our modelled SO$_2$ abundance at 4~\rs\ is $\sim 4 \times 10^{-8}$, a value consistent with the T Cep observations. Decin et al. (2010a) derive a high abundance of $1\times 10^{-6}$ from fitting emission lines originating in the wind between $50$~\rs\ and $200$~\rs. This region is located outside the inner wind and overlaps with the outer envelope, defined as the region where the interstellar ultraviolet  radiation field starts dissociating molecules and triggers a cold, ion-rich chemistry. Combined with the penetration of cosmic rays, the region starts around 100~\rs. The high abundance of SO$_2$ derived from observations may not reflect a high molecular abundance in the inner wind at $r< 10$~\rs, but a possible high molecular abundance at the onset of the outer envelope. 

The region between $~10$~\rs\ and $100$~\rs, labelled ``the intermediate envelope'' in Figure~\ref{fig1}, is characterised by cool gas temperatures ($\sim 100$ K), and the presence of parent molecules and dust formed in the inner wind. It has long been hypothesised that some molecules commonly detected in AGB winds, e.g., methane, CH$_4$ in carbon stars and ammonia, NH$_3$, in AGBs, could originate from this region (\cite{om88}). Indeed, for IK Tau, we form very small amounts of NH$_3$ from the gas phase in the inner wind, as seen from Table~\ref{tab3}, while a large abundance has been derived from recent Herschel observations (\cite{men10}). These data indicate a formation region with gas number densities of $\sim 10^6$ \cmc\ and gas temperatures in the range $10-100$K, typical of the intermediate envelope. Therefore, molecules that cannot form in the shocked inner wind from the gas phase could well be synthesised on the surface of dust grains in the intermediate envelope.  

A different scenario applies to PO. The phosphorous chemistry is very poorly characterised and we have simply used the isovalence between nitrogen and phosphorous to derive some chemical rates for P-bearing species. While the modelled abundance for PN slightly exceeds but agrees well with the observed value, the modelled PO abundance is $\sim 3$ orders of magnitude smaller than the value derived from observations by de Beck et al. (2013). In the phosphorous chemistry, the dominant formation process for PN is the reaction
\begin{equation}
\label{eq11}
{\rm PO+ N \rightarrow PN + O,} 
\end{equation}
while PO essentially forms from the reaction
\begin{equation}
\label{eq12}
{\rm P + OH \rightarrow PO + H, }
\end{equation}
and 
\begin{equation}
\label{eq13}
{\rm PN + OH \rightarrow PO + NH.}
\end{equation}
The reaction rates of reactions \ref{eq11}$-$\ref{eq13} have been estimated from the N/P isovalence and have not been measured or theoretically assessed. However, the formation of PO is linked to that of PN. Then, the slight overestimate of the depletion of P atoms in PN as shown by our modelled value may result in lowering the PO abundance. We thus attribute the discrepancy between modelled and observed PO abundances to a poorly characterised chemistry but note that the isovalence principal gives a satisfactory match for PN. 

\begin{figure}
\centering
  \includegraphics[width=9.2cm]{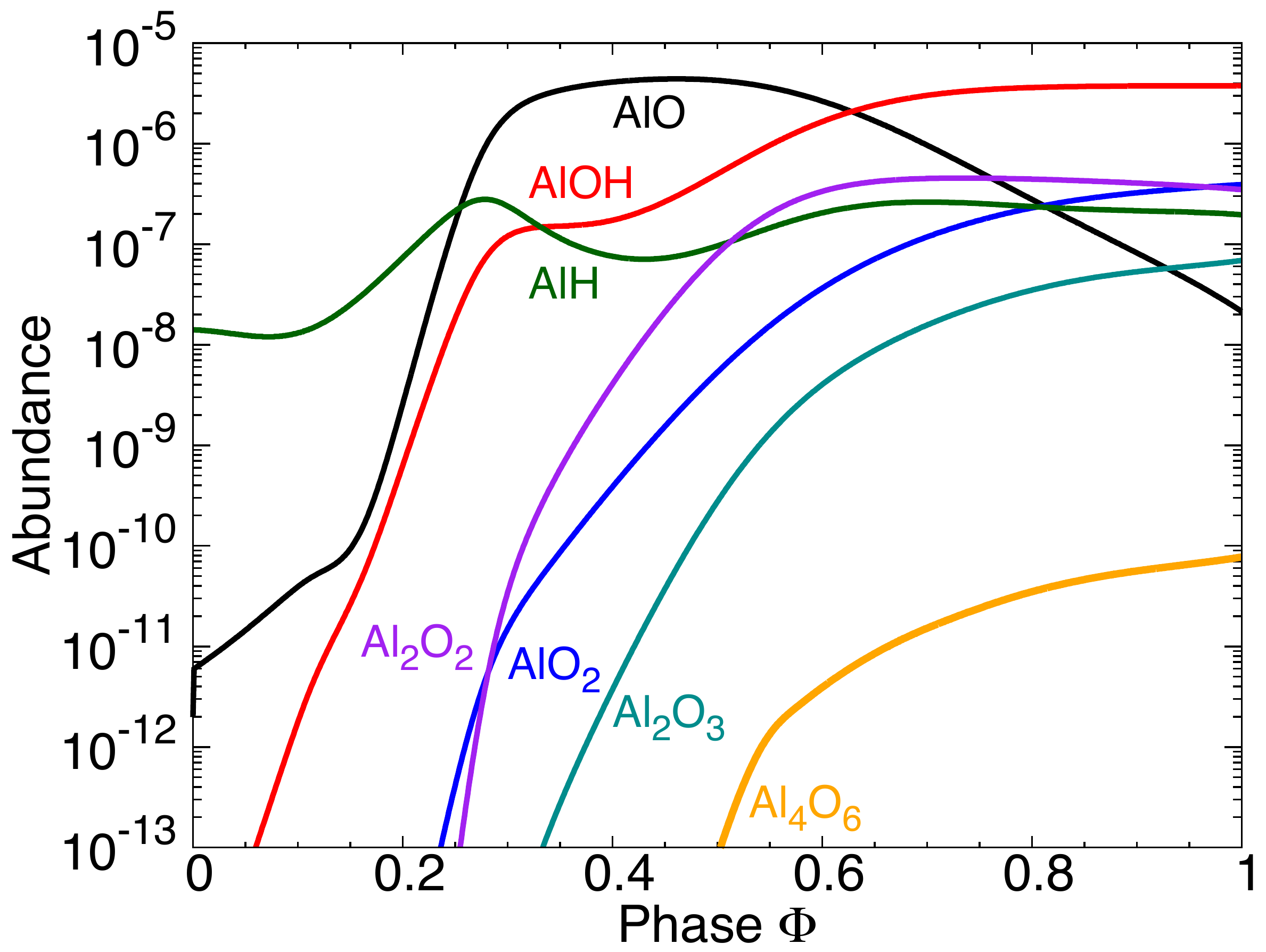}
\caption{Modelled abundances of Al-bearing species and \al\ clusters at $r = 1$~\rs.}
\label{fig6}
\end{figure}

\subsection{Dust}
\label{dust}
The formation of the solid phase is coupled to the gas-phase chemistry and calculated at each integration time step in the post-shock gas. We present results on grain size distributions for both alumina and silicate grains, and list the derived dust-to-gas mass ratio in Table \ref{tab6}. The build-up of dust grains occurs over a number of pulsations that are necessary for the gas layer to move from $r$~\rs\ to ($r + \delta r$)~\rs. We calculate this number of pulsations by assuming a local gas drift velocity. The drift velocity is small close to the star owing to stellar gravity, and increases further away from the stellar surface. We choose drift velocity values derived from pulsation models, and assume a drift velocity of 0.5 \kms\ for $r < 2$~\rs\, and of 1.5 \kms\ for $r > 3$~\rs. 
\begin{figure}
\centering
  \includegraphics[width=9.2cm]{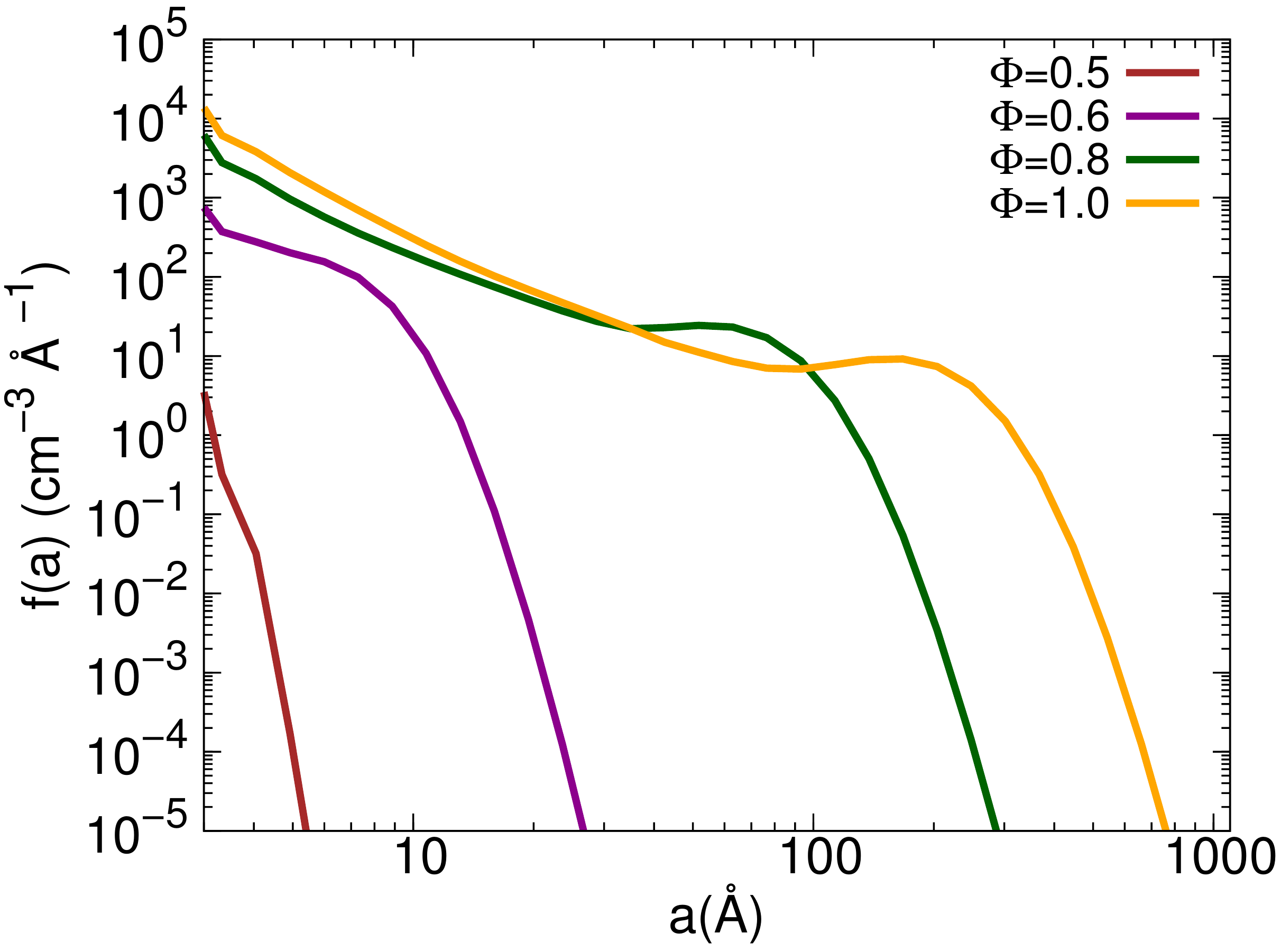}
  \includegraphics[width=9.42cm]{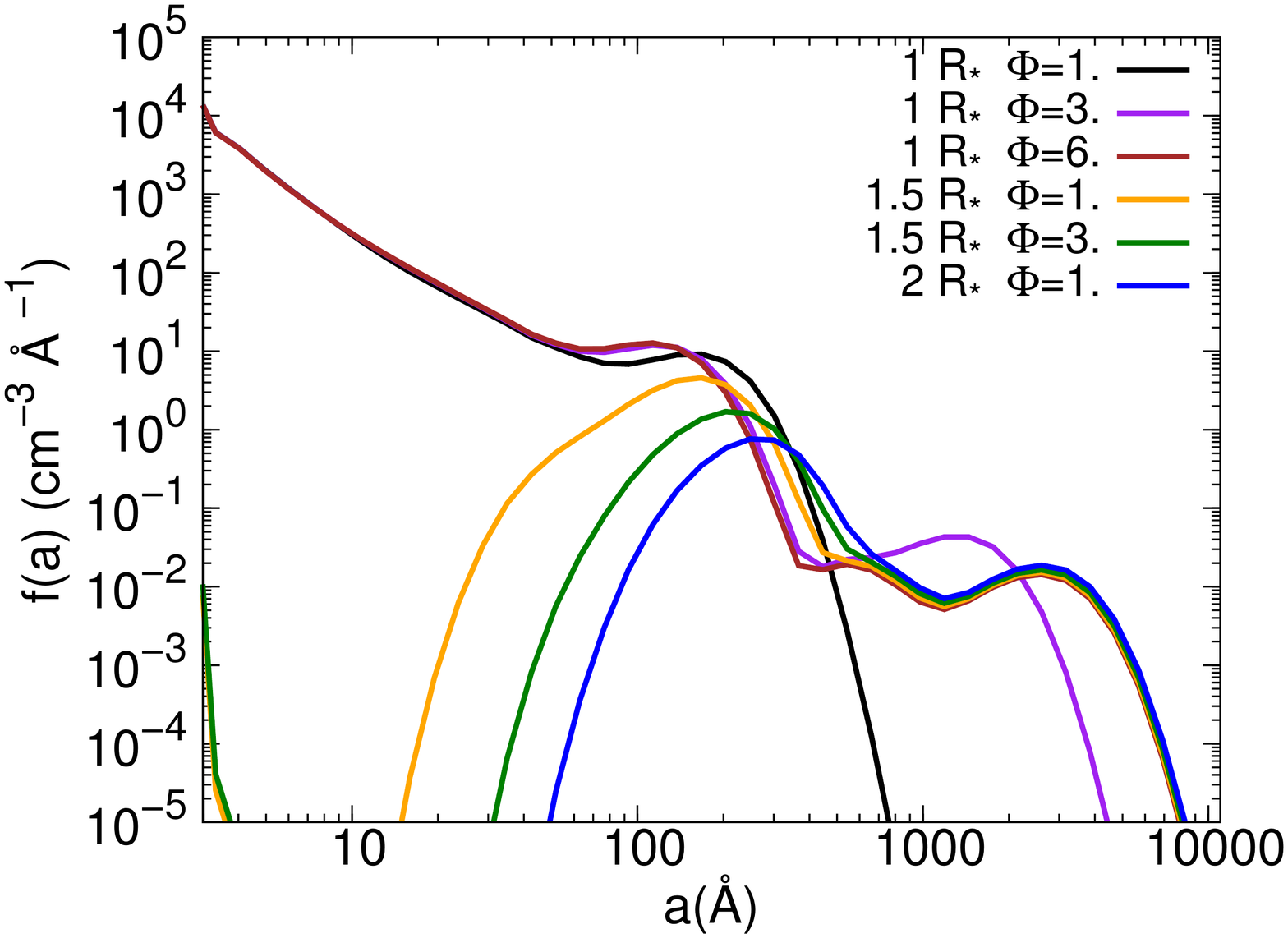}
\caption{Grain size distributions for alumina dust formed in the shocked gas layer. Top: as a function of phase in one oscillation over one pulsation period at 1~\rs; Bottom: the shocked layer has moved from 1~\rs\ to 2~\rs. Distributions are shown for different pulsation phases $\phi$ (with $\phi=1-6$) at radius $r = 1-2$~\rs.}
\label{fig7}
\end{figure}

Owing to the high gas temperatures characterising the post-shock gas at 1~\rs\ (see Figure \ref{fig2}), it is reasonable to consider that the dust clusters and small grains that form in one oscillation at 1\rs\ may not survive the passage of the next shock after one pulsation period. Any dust formed at 1\rs\ may thus not accumulate and grow over several pulsation periods in the shocked photosphere. The situation is different at larger radii, where the post-shock gas temperature drastically decreases.

\subsubsection{Alumina}
\label{alu}

The alumina clusters chiefly form in the post-shock gas at 1~\rs, and their abundances in the first oscillation are illustrated in Figure \ref{fig6}. The formation of \al\  involves the dimerisation of AlO through termolecular recombination while AlO forms from the reaction of Al with OH in the first place. The dimerisation of \al\ is also described by termolecular recombination to form the \al\ dimer, (\al)$_2$, which has a binding energy of $\sim 15.7$~eV (\cite{li12}), and can thus withstand the high temperatures of the post-shock gas. This dimer is considered as the seed that will further condense into alumina grains. Over one oscillation resulting from the first shock at 1~\rs, only a small quantity of alumina dimers forms, reaching an abundance of $1\times 10^{-10}$ at $\phi=1$. A large fraction of Al stays in the form of molecular rings (e.g., Al$_2$O$_2$ and Al$_2$O$_3$). Because the dimerisation of AlO and \al\ is termolecular and thus dependent on the background pressure, cluster formation occurs over a reduced range of radii in the inner wind and the formation of \al\ cluster ceases after 2~\rs. 

The grain size distributions for alumina in one oscillation at 1~\rs\ is presented in Figure \ref{fig7} as a function of pulsation phase. Because of the high post-shock gas temperatures after the shock passage and the extremely low abundance of alumina dimers, no grains form at $0 < \phi <0.5$. Very small grains start forming after $\phi=0.5$, but coagulation starts to be efficient for dust growth in the phase range $\phi=0.8-1$, for which gas temperatures span the range $\sim 1600-2000$ K and the alumina cluster abundance reaches a maximum. The distribution peaks at grains in the size range $70-200$ \AA, despite the small fraction of \al\ clusters available. However, the post-shock gas densities are very high at 1~\rs, as seen from Table \ref{tab1}, and the clusters coagulate and condense with high efficiency and lead to the formation of \al\ grains of small size over one oscillation and pulsation period. Under the assumption that small alumina grains may not survive the passage of periodic shocks, the results show that there exists a steady population of small size ($a \sim 0.02$ \mic) alumina grains located in the shocked photosphere at 1~\rs. If we consider that grain growth may also proceed through surface deposition when the grain radius has reached several tens of \AA\ (at $\phi=0.8$ and after), the grain size will increase, but the growth will be time-limited and proceed only between $\sim 0.8 \le \phi \le 1$. 

We now consider the extreme case where alumina grains survive the periodic shocks at each shock passage, while molecules are destroyed and reform. We may then investigate grain growth over several pulsation periods. For a gas drift velocity of 0.5~\kms, the gas layer at 1~\rs\ takes 12 pulsations to reach 2~\rs. We thus consider the build-up of dust grains over a number of 12 pulsations for the standard gas conditions given in Table \ref{tab2}, and show the resulting alumina grain size distributions in Figure \ref{fig7}. The grain size distributions are skewed towards large grain sizes. At 2~\rs, the distribution comprises two dust grain components peaking at $\sim$ 300~\AA\ and 0.3~\mic. All small grains produced in the first oscillation at the shocked photosphere have effectively condensed in larger grains after 12 pulsations. The distribution does not change at $r > 2$~\rs, owing to a shortage of growing agents as all alumina clusters have been included in dust grains. The mass of alumina dust produced remains modest and the dust-to-gas mass ratio varies from $3.3\times 10^{-6}$ at 1~\rs\ over one oscillation to $2.8\times 10^{-5}$ at 2~\rs\ when grain survival is assumed, as seen from Table~\ref{tab6}. 

\begin{table*}
\caption{Derived dust-to-gas mass ratio for alumina and silicate dust grains as a function of radius for the gas number densities given in Table \ref{tab2} and the gas number densities increased by a factor of 10 (labelled``x10''). The radius range where strong wind acceleration starts is indicated in italic. }           
\label{tab6}       
\centering                         
\begin{tabular}{l c c c c c c}        
\hline\hline      
$r$ (\rs) & \multicolumn{4}{c}{Alumina}&\multicolumn{2}{c}{Silicate}\\
\hline 
 & 1 Osc.& Accumulated & 1 Osc. (x10) & Accumulated (x10)& Accumulated & Accumulated (x10)  \\    
 \hline     
1.0 & $ 3.3 \times 10^{-6}$ &$ 2.1 \times 10^{-5}$ &$ 6.7 \times 10^{-5}$ & $ 4.3 \times 10^{-4}$& --& --\\
1.5 & $ 2.6 \times 10^{-16}$ &$2.7 \times 10^{-5}$ &$ 5.4 \times 10^{-9}$ &$ 5.3 \times 10^{-4}$ &-- &-- \\
2.0 & $6.7 \times 10^{-27}$ & $ 3.3 \times 10^{-5}$& $ 3.6 \times 10^{-17}$ &$ 6.5 \times 10^{-4}$ &-- & --\\
3.5 &'' & "& "& "&$ 1.3 \times 10^{-4}$& $ 4.7 \times 10^{-4}$\\
4.0 &" & "& "& "&$ 4.6 \times 10^{-4}$&$ 1.1 \times 10^{-3}$ \\
5.0  &" & "& "&"&$ 7.3 \times 10^{-4}$& $ 3.0\times 10^{-3}$\\
{\it 6.0 } &" & "& "&"&${ \it 1.0 \times 10^{-3}}$&$  {\it 5.1 \times 10^{-3}}$\\
{\it 7.0} &" & "& "&"&$ {\it 1.4 \times 10^{-3}}$&$ {\it 6.1 \times 10^{-3}}$ \\
{\it 8.0} &" & "& "&"&$ {\it 1.7 \times 10^{-3}}$&$ {\it 6.7 \times 10^{-3}}$ \\
{\it 9.0} &" & "& "&"&$ {\it 2.0 \times 10^{-3}}$&$ {\it 7.5 \times 10^{-3}}$ \\
{\it 10.0} &" & "& "&"&$ {\it 2.3 \times 10^{-3}}$&$ {\it 8.0 \times 10^{-3}}$ \\
\hline                      
\end{tabular}
\tablefoot{``1 Osc.'' corresponds to the case where the dust forms over one oscillation but is destroyed in the next shock.``Accumulated'' corresponds to the case where the dust forms over one oscillation, is not destroyed by the next shock and accumulates over several pulsation periods - see text.}
\end{table*}

\subsubsection{Silicates}
\label{sili}

\begin{figure}
\centering
\includegraphics[width=\columnwidth]{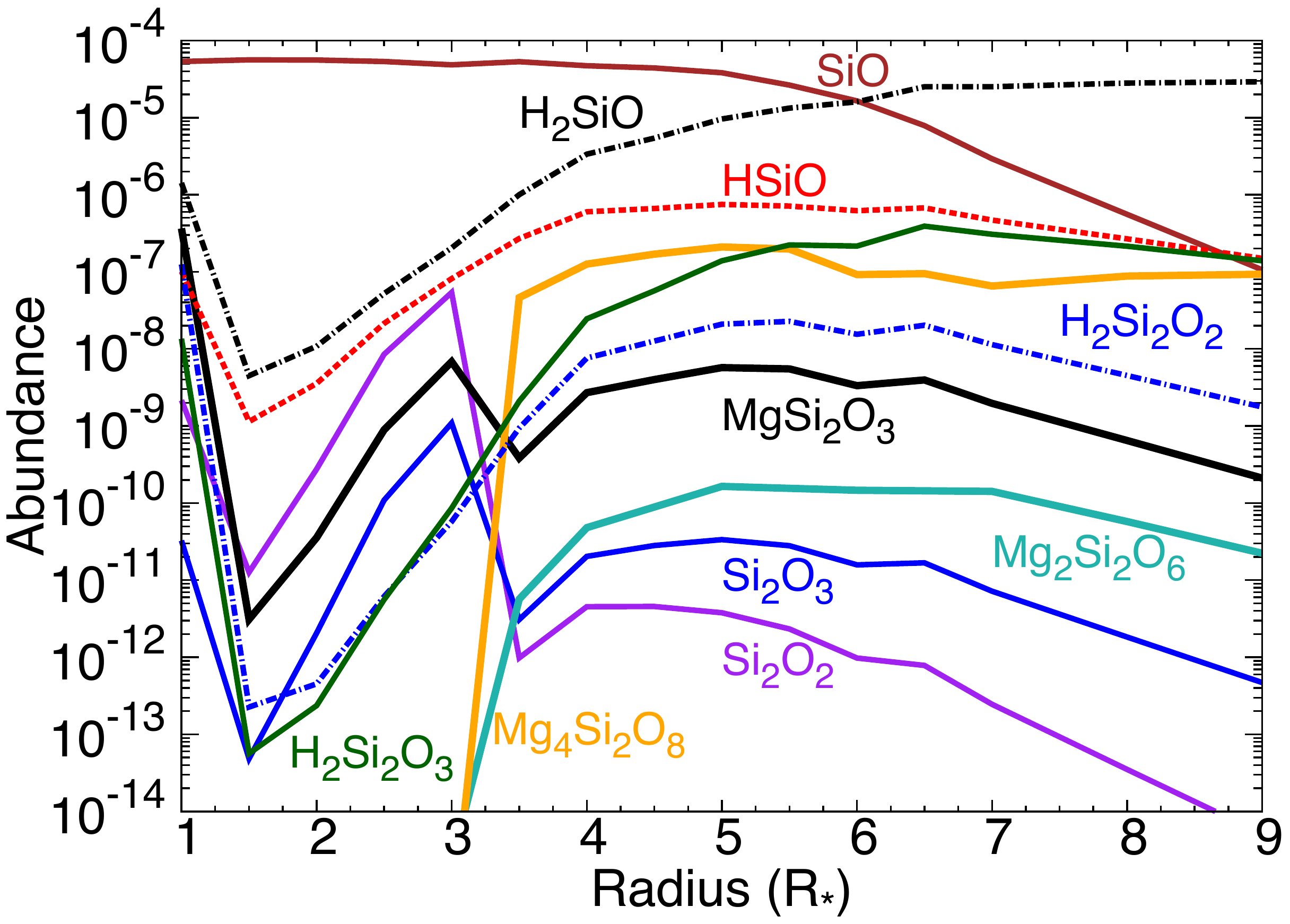}
\includegraphics[width=9.1cm, height=6.65cm]{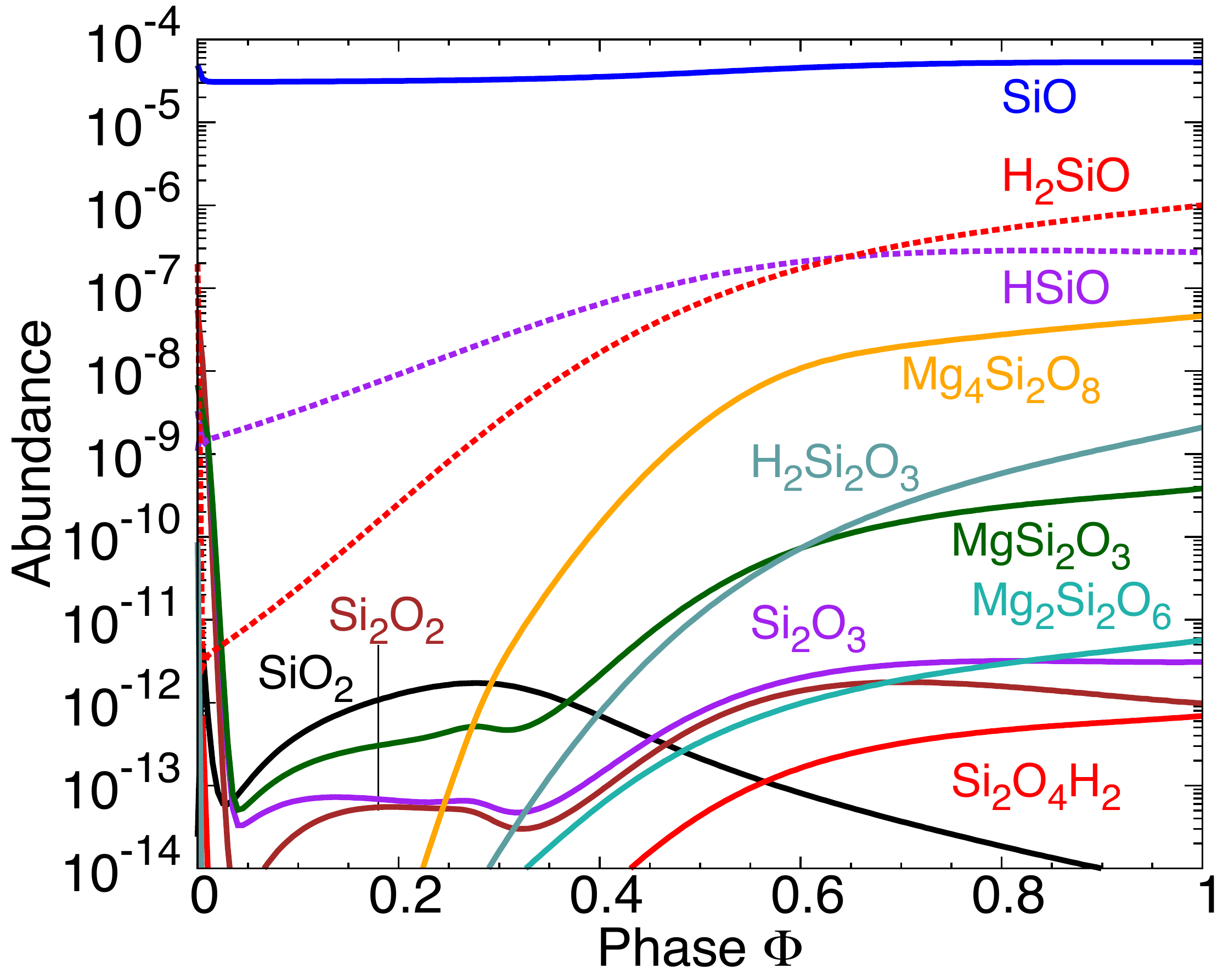}
\caption{Top: Abundances with respect to total gas number density n$_{gas}$ of the silicate clusters and key molecules as a function of radius; Bottom: Abundances with respect to total gas number density n$_{gas}$ of the silicate clusters as a function of pulsation phase at $r=3.5$~\rs.}
\label{fig8}
\end{figure}
Small clusters of silicates form further out of the shocked, stellar photosphere, as seen in Figure \ref{fig8}, where abundances of key molecules involved in the nucleation of small silicate clusters are shown as a function of radius. The silaformyl radical, HSiO, and silanone, H$_2$SiO, start to efficiently form at $r > 1.5$~\rs\, because the reverse processes of reactions R1--R4 in Table \ref{tab4} are favoured and the destruction of these nucleating species ensues at shorter radii. The synthesis of dimers of forsterite stoichiometry starts at $r > 3$~\rs, and the abundances of the prevalent clusters involved in the nucleation phase are shown in Figure \ref{fig8} for the gas at $r=3.5$~\rs. The gas temperature at which all intermediate clusters are effectively converted into silicate dimers is $\sim 1200$~K in the post-shock gas, where the gas number density is $\sim 4.5\times 10^{10}$~\cmc. 

\begin{figure}
\centering
  \includegraphics[width=\columnwidth]{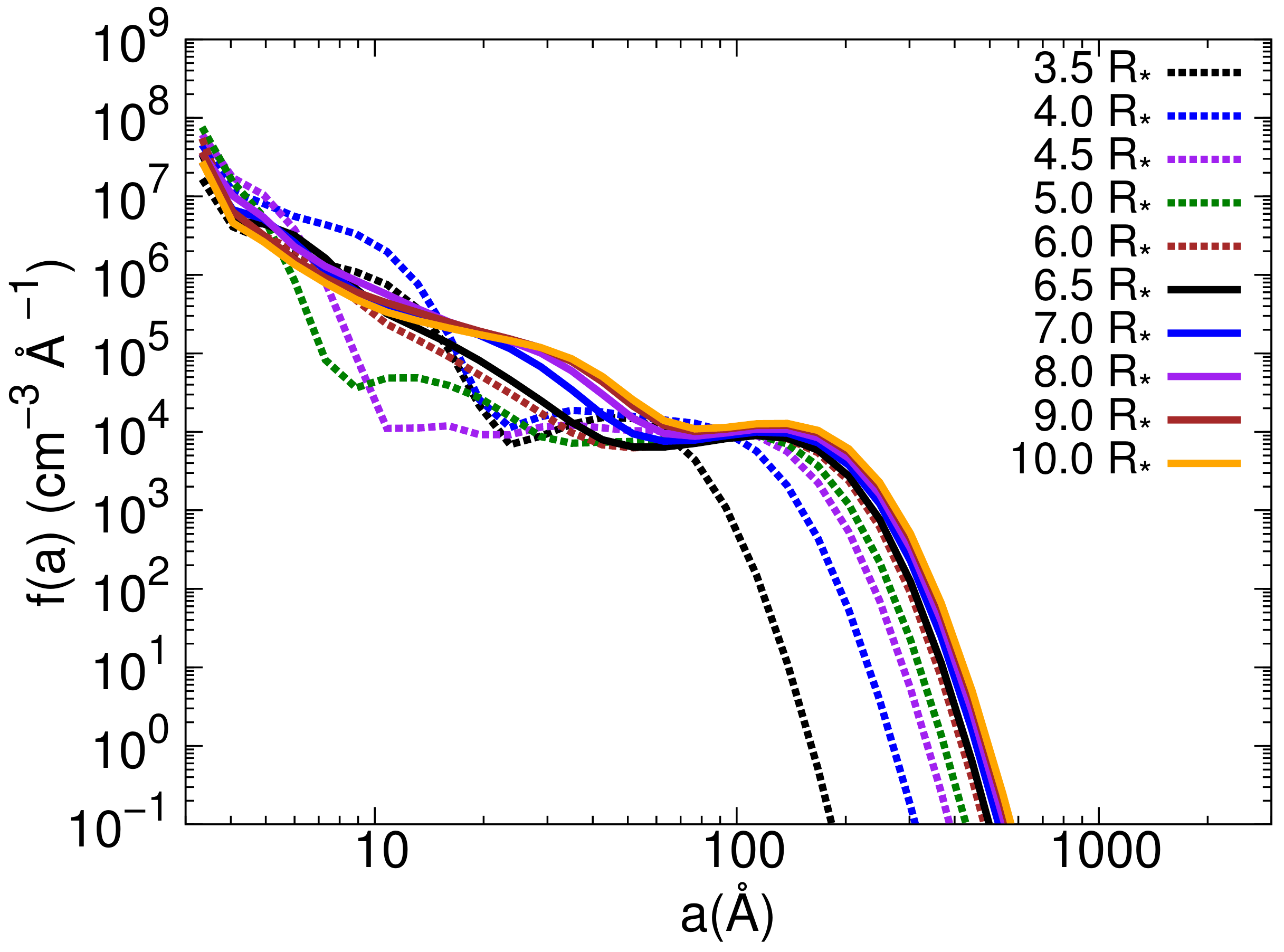}
\caption{Grain size distributions for silicate dust formed in a shocked gas layer that has moved from 3.5~\rs\ to 10~\rs.}
\label{fig9}
\end{figure}

The build up of silicate dust takes place from 3.5~\rs\ to larger radii, over several stellar pulsations, as we assume that the small silicate grains formed in one oscillation at 3.5~\rs\ withstand the next shock passage. As mentioned in \S\ \ref{dust}, for a gas drift velocity of 1.5~\kms, it takes 2 pulsations to travel a distance of 0.5~\rs. Therefore, 10 pulsations are necessary to move from 3.5~\rs\ to 6~\rs, where the growth of silicate grains slows down and almost reach completeness at 8~\rs. Figure~\ref{fig9} shows the silicate grain size distributions at various radii and pulsation numbers. We see that the size distribution has a tail of large grains that peaks around $50$ \AA\ at 3.5~\rs\ and reaches $\sim 200$ \AA\ at $r \ge 6$~\rs. The dust-to-gas mass ratio listed in Table \ref{tab6} reaches $ 1.0 \times 10^{-3}$ at 6~\rs, and the ratio keeps increasing out to 10~\rs, where it reaches $ 2.3 \times 10^{-3}$. These values agree with dust-to-gas mass ratios derived from observations of O-rich Mira stars. Based on the derived gas-to-dust mass ratios, we estimate that  $\sim 22$~\% of the elemental silicon abundance is depleted in silicate grains at 6~\rs. 

Because the growth of silicate grains takes place between 3~\rs\ and 6~\rs\ and slows down starting at 6~\rs, the wind acceleration, which is triggered by radiation pressure on dust grains, should start developing as soon as enough dust grains form in the radius range $3.5-4$~\rs, and increase to reach full regime when the grain sizes are large enough. As for alumina, the deposition of chemical species on the grain surface will also proceed along with coalescence and coagulation as soon as the grain radius exceeds several tens of \r{A}ngstr{\"o}ms. This may happen as near as $3.5-4$~\rs\ as shown in Figure~\ref{fig9}, where some dust grains reach a radius of $\sim 50$~\AA\ at 3.5~\rs\ and $\sim 100$~\AA\ at 4~\rs. We expect that deposition will push the final grain size distribution to radii $a > 200$~\AA, and increase the final dust-to-gas mass ratio, but cannot yet quantify the increase. From Figure \ref{fig5}, we note that the modelled SiO abundance derived at 6~\rs\ is $\sim$ twice as large as the value derived from observations. The modelled SiO value is then consistent with the potential growth of dust grains through gas deposition along with ongoing coalescence and coagulation. The prevalent wind acceleration zone may then encompass the radius range $3.5-6$~\rs. This acceleration will result in the freezing of molecular abundances after $r > 6 $~\rs. This is the underlying reason for comparing modelled abundances at $6-8$~\rs\ with values derived from astronomical data in \S~\ref{gas} and Figure~\ref{fig5}. 


\subsubsection{A high gas number density case}

Despite the fact that the present standard model for IK Tau provides satisfactory dust-to-gas mass ratios for silicates compared to values derived from observations, the grains in which most of the silicate mass is concentrated are of medium size (i.e., $\sim 200$ \AA). We thus test the impact of the gas number density on the synthesis of clusters and dust grains by increasing the gas number density at the photosphere by a factor of ten and keeping all photospheric parameters and the shock characterisation as before. This increment in number density is certainly too high to properly describe the IK Tau wind at the base of the intermediate envelope, defined as where the wind is fully accelerated. Assuming the mass loss rate and wind terminal velocity given in Table~\ref{tab1}, we derive from the definition of mass loss rate and conservation of mass,

\begin{equation}
\label{eq16}
n(r)= {\dot{M} \over{4\pi m_H \mu_{gas} v_{exp}  r^2}} = {1.09 \times 10^{10} \over {r'^2}},
\end{equation}
where $m_H$ is the mass of the hydrogen atom, $\mu_{gas}$ the gas mean molecular weight at the acceleration radius (taken as 2 a.m.u. because the gas is essentially in the form of H$_2$), $v_{exp}$ the terminal wind velocity, and $r'$ the position in the wind in units of stellar radius~\rs. From equation \ref{eq16}, we deduce a gas number density of $\sim 1.7 \times 10^8$ \cmc\ at $r= 8$~\rs, a value that is compatible with the number density of the pre-shock gas at 8~\rs\ given in Table \ref{tab2}. Therefore, our modelled wind parameters are consistent with a wind acceleration fully complete at $r \sim 8$~\rs. This radius corresponds to the position at which the synthesis of silicate grains has ended. 

An increment of $10$ in the gas number density pushes the radius at which the wind is fully accelerated to $r > 20$~\rs, according to equation \ref{eq16}, and does not represent a model for the dust formation zone. However, such a number density enhancement may indicate trends produced by the presence of over-dense regions in the inner wind. 

\begin{figure}
\centering
  \includegraphics[width=9.18cm]{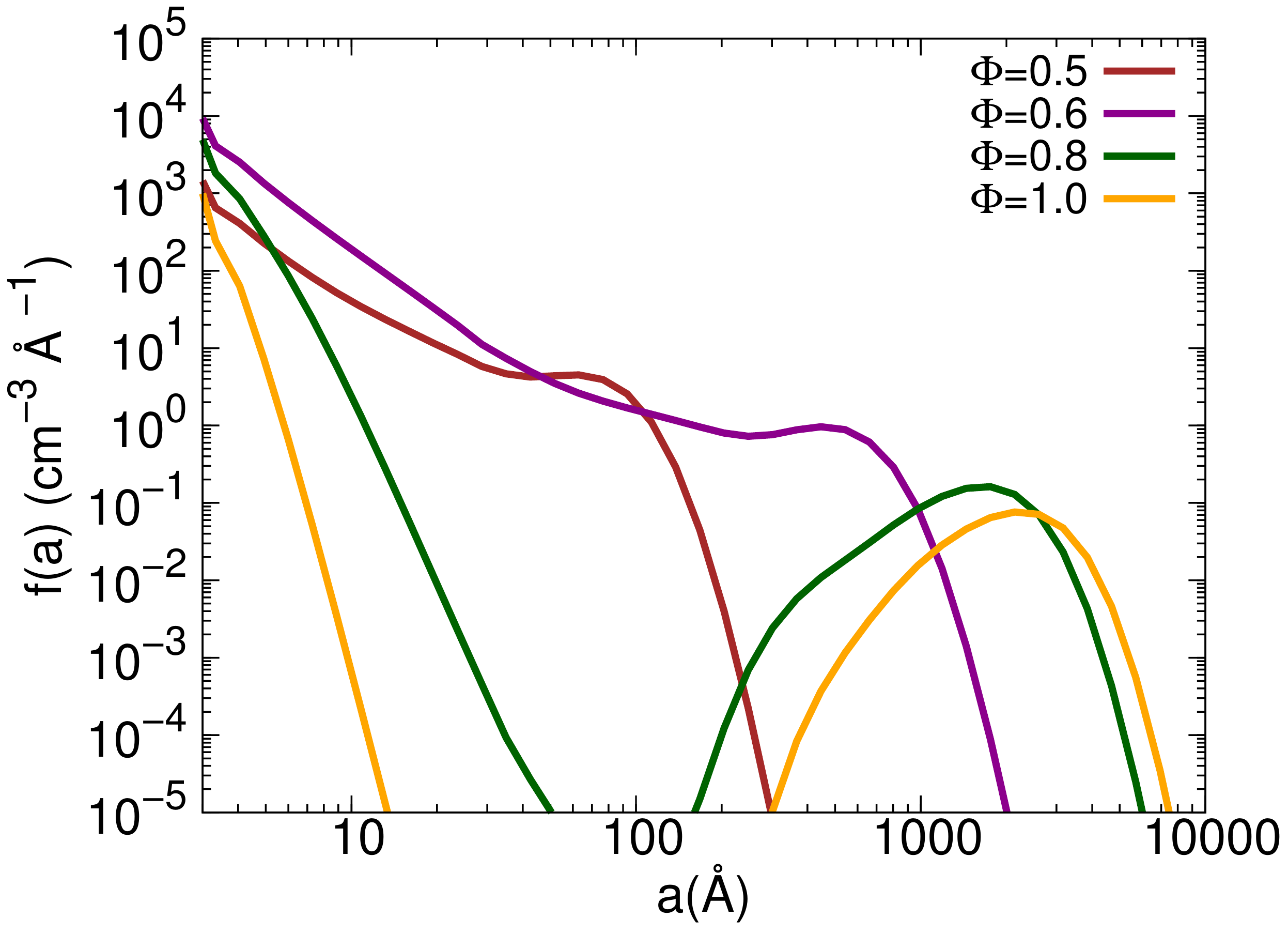}
  \includegraphics[width=\columnwidth]{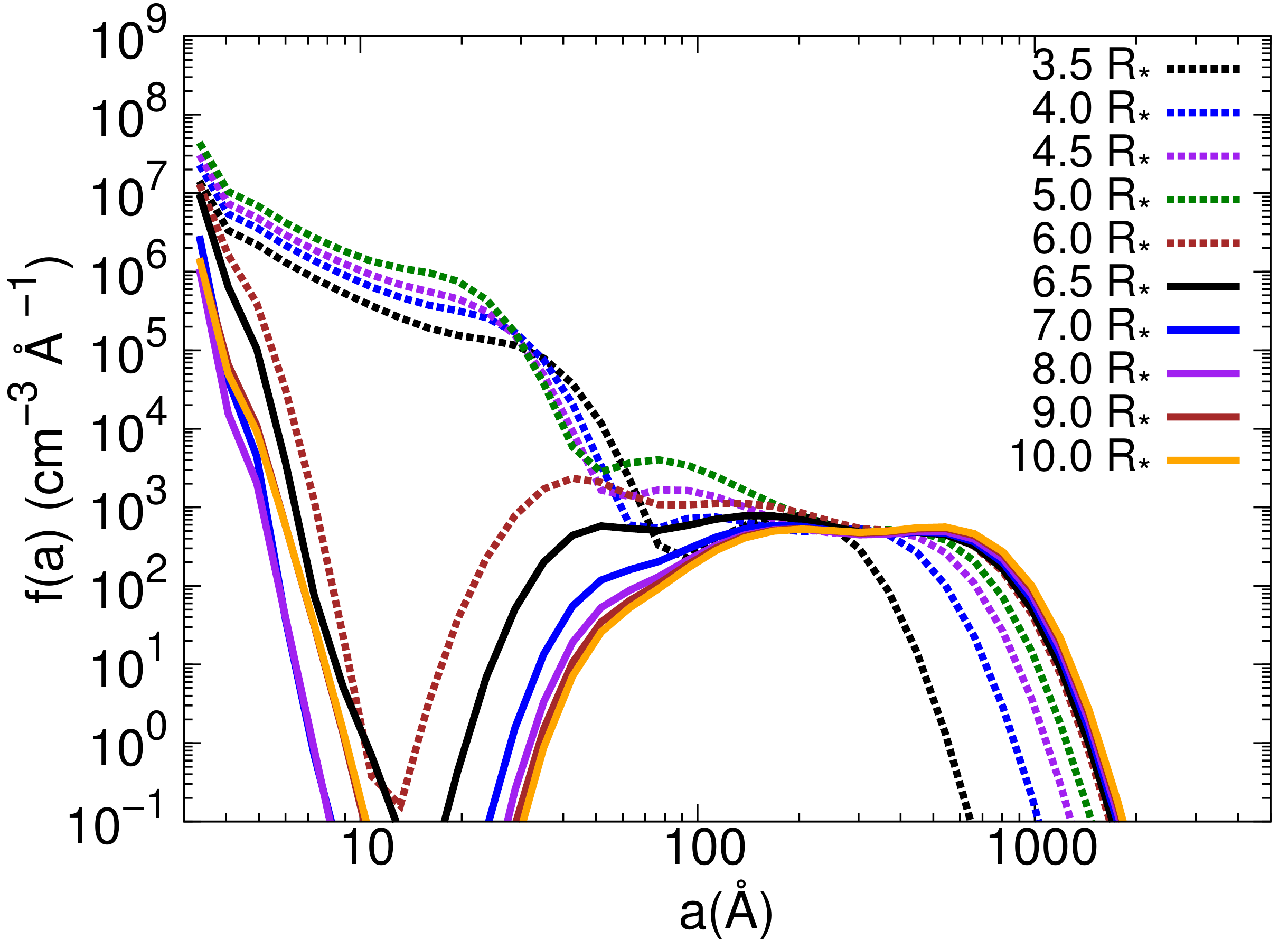}
\caption{Grain size distributions for  a photospheric number density x10. Top: alumina in a shocked gas layer that has moved from 1~\rs\ to 2~\rs; Bottom: silicate in a shocked gas layer that has moved from 3.5~\rs\ to 10~\rs.}
\label{fig10}
\end{figure}

In Figure \ref{fig10}, the grain size distributions for alumina at 1\rs~and silicates at $r = 6$~\rs\ are shown. We see that a denser gas produces much larger grains. For alumina, we consider the formation over one oscillation and no grain growth over several pulsations. We see that most of the dust mass resides in large grains of size $a \sim 0.2$ \mic, while the size distribution peaks at $a \sim 200$ \AA\ when no density enhancement is assumed (see Figure~\ref{fig7}). If deposition on grain surface were considered, grains could still grow between $\phi = 0.5 -1$, to possibly start triggering a wind through Mie scattering very close to the star (\cite{hof08}). 

As for silicates, the grains reach an average size of $\sim 0.1$~\mic\ at $r \ge 6$~\rs, with a dust-to-gas mass ratio of $5 \times 10^{-3}$ at 6~\rs. Because silicates form at larger radii than alumina, the impact of increasing the gas number density on the final grain size is weaker than for alumina, which forms in very dense layers at $r \le 2$~\rs. However, any inhomogeneity in the inner wind will result in the efficient synthesis of a population of large grains for both materials, and contribute to gas drag and wind acceleration close to the star.

\section{Summary and discussion}
\label{sec5}

Expanding the previous study by Duari et al. (1999), we present a comprehensive model of the inner wind of the M-type AGB star IK Tau,  where the chemistry of the gas phase and dust clusters is coupled to the dust condensation phase, thereby providing a self-consistent treatment of the final dust masses and grain size distributions. The main results are as follows:

\begin{itemize}
\item{The modelled abundances for the molecules so far detected in the inner wind of IK Tau agree well with values derived from observations, except for PO and SO$_2$ (for the latter and NH$_3$, the formation possibly occurs in a region located outside the inner wind at $r > 10$~\rs). While some molecules, already present in the photosphere, are destroyed by the shocks and reform in the inner wind with abundances close to TE values, other species are only produced in the inner wind by shock-induced chemistry. The agreement for most molecular species indicates that the periodic passage of shocks in the inner regions close to the star has a strong bearing on shaping the upper atmosphere and triggering the synthesis of molecules by non-equilibrium chemistry in AGB stars, as already suggested by Willacy \& Cherchneff (1998), Duari et al. (1999) and Cherchneff (2016, 2011, 2012).}
\item{The prevalent gas-phase dust clusters in the nucleation phase are alumina and magnesium-rich silicates of forsterite stoichiometry (olivine). Enstatite (pyroxene) clusters are not abundant, with abundance values at least three orders of magnitude lower compared to forsterite. Small clusters of metal oxides (MgO and FeO) do form with low abundances. As for alumina, the clusters form very close to the star at $r \le 2$~\rs, while silicate clusters form at larger radii ($r \ge 4$~\rs).}
\item{For our standard model of the IK Tau inner wind, one population of small alumina grains forms close to the star, peaking at a grain size of $\sim 200$ \AA. This population might be destroyed at each shock passage but will reform at a later phase in the  pulsation. Therefore, our results indicate that a steady population of small alumina grains forms close to the star. The silicate population includes a component of large grains with a distribution that peaks at $\sim 180$ \AA\ after 6~\rs. The mass of silicates is distributed in grains of smaller sizes because of the lower gas number density at which the condensation takes place compared to alumina. When we consider an enhanced gas number density at the stellar photosphere, the size distributions of alumina and silicates are skewed towards larger grains, where alumina grains reach $\sim 0.2$~\mic\ in size and silicate grains peak around 0.08 \mic. Since we do not model the growth of grains through surface addition, these results can be taken as lower limits for the grain size and mass of alumina and silicates.} 
\item{The total dust-to-gas mass ratios derived at 6~\rs\ range from $\sim 1 \times 10^{-3}$ and $6 \times 10^{-3}$ for the standard and the enhanced density cases, respectively. This is in good agreement with dust-to-gas mass ratios derived from observations for IK Tau. }
\end{itemize} 

Several recent interferometric studies of oxygen-rich AGB stars point to the presence of an alumina shell close to the star. While Norris et al. (2012) found a halo of large ($\sim 0.3 $ \mic), transparent grains close to the photosphere (at $r \le 2$~\rs) in a few O-rich AGBs, Zhao-Geisler et al. (2012) reported evidence for the coexistence of a warm layer of H$_2$O and a \al\ dust shell close to the stellar surface in several O-rich evolved stars. The presence of two dust shells of alumina and silicate composition, where the alumina shell is located below the silicate shell, have been detected in a few O-rich stars (\cite{zhao12, kar13}).  This latter interferometric study points to the formation of alumina dust at $\sim 2$~\rs\ and of silicates at $\sim 5$~\rs\ from the stellar surface for the few sources showing the typical alumina and silicate features in the mid-infrared. Therefore, these various studies point to a dust condensation scenario similar to that derived from the present study, with possibly two separate dust populations, one close to the star and consisting of alumina grains, the other further out and made of silicates.  

Owing to the small opacities of iron-free silicates in the wavelength range where most of the stellar emission flux is produced (around 1 \mic), acceleration of the gas by photon scattering on large forsterite grains was proposed as a substitution to photon absorption (\cite{hof08}, \cite{bla12}, 2013). According to these studies, Mie scattering on grains in the size range $0.1-1$ \mic\ is able to generate a wind acceleration and terminal velocity, and mass loss rates that agree with values derived from observations. Whether such large forsterite grains really form in the wind requires a careful chemical kinetic-based investigation, because models by H{\"o}fner (2008) and Bladh et al. (2012, 2013) do not treat the nucleation of forsterite grains but describe growth by assuming a pre-existing populations of small seeds in the inner wind. The interferometric study of a few oxygen-rich AGBs by Norris et al. (2012) indicates scattered light dust shells very close to the star, at radii $\le 2$~\rs. An iron-free silicate grain composition is favoured because iron-rich silicates would be too hot to survive so close to the star owing to their high near-IR opacity, but alumina grains are not excluded and can equally well reproduce the observations. The derived average grain sizes are large for all objects, and around 0.3 \mic. Our chemistry-based model of IK Tau shows that silicates of any type should not form at radius $ < 3.5 $~\rs, because the formation of small molecular clusters involved in the nucleation of silicates is hampered. On the other hand, alumina dust clusters can easily form close to the star from simple three-body association of AlO molecules, and further oxidation to form the small ring \al, and the subsequent alumina dimers at gas temperatures less than 2500 K. The high gas density warrants a rapid and efficient condensation of these alumina clusters to form grains, whose total mass is distributed in a population of medium-size  grains with radius $\sim 0.02$ \mic. Possible surface growth will take place when the post-shock temperature is low enough to allow the grain survival and growth, which will result in larger grains. These results and trends then support the presence of an alumina shell very close to the stellar surface, as observed by Norris et al. (2012).

The formation of silicates takes place at larger radii ($r > 3.5 $~\rs) because the nucleation scheme involves the formation of several molecular intermediates, specifically the silaformyl radical, HSiO, and silanone, H$_2$SiO, whose dimers cannot form close to the star. Our model for IK tau gives silicate grains of medium size in the range 200 \AA\ $-$ 0.1 \mic, and most of our forsterite grains are formed between 6~\rs\ and 8~\rs, at which point condensation stops because the local gas density is too low to facilitate efficient coalescence and coagulation. In this radius range, the wind will be accelerated and this process will gradually freeze the gas-phase chemistry and the formation of molecules and dust clusters. It is thus unlikely that the dust shells derived from interferometric studies, and found close to the stellar surface, consist of iron-free silicates (e.g., \cite{sac13}). In our model, grains of alumina are formed close to the star. Because they are likely destroyed at each shock passage and reform in the post-shock gas, they will reside close to their formation location without being expelled at larger radius. If so, and as mentioned before, our results indicate the presence of two distinct grain populations in the inner wind, as recently detected by Karovicova et al. (2013). We note that the various interferometric observations often relate to semi-regular objects and not variable Mira-type stars like IK Tau. Because semi-regular AGB stars usually have large photospheric temperatures, the formation of silicates may be delayed to even larger radii, and thus much smaller amounts of small silicate grains will form under more diffuse gas conditions. Therefore, alumina might be the major dust component in semi-regular variables whereas silicate dust might become the prevalent dust type in the cooler and more evolved oxygen-rich AGB stars. 

As already mentioned, we do not consider the growth of dust grains through surface addition of gas-phase species. It is often assumed that some small seeds formed close to the stellar surface could act as condensation nuclei and grow from heterogeneous addition of material on the surface. Specifically, TiO$_2$ has often been proposed as a condensation seed in AGB winds and brown dwarf clouds because of its refractory nature (\cite{jeo03,lee15}). Furthermore, Goumans \& Bromley (2013) showed that one TiO$_2$ molecule was enough to kick-start silicate dust formation via SiTiO$_3$ seeds, thus overcoming the SiO dimerisation bottleneck. Our results on gas-phase TiO/TiO$_2$ indicate that titanium remains in atomic form and in TiO in the inner wind, with some TiO converted into TiO$_2$ by shock chemistry. Several transitions of molecular TiO and TiO$_2$ have been observed in the wind of the supergiant VY~CMa (\cite{kam13a, deb15}), where the TiO$_2$ emission is extended and not always correlated with the dust continuum emission. These results point out that TiO$_2$ clusters and grains may not play an important role as condensation seeds. It seems that the modest amount of TiO$_2$ we derive in the gas phase precludes solid TiO$_2$ to be a key agent in the dust formation process in IK Tau and other O-rich AGBs. This conclusion was also drawn by Plane (2013) in his study of calcium titanate formation in AGB stars. 

The formation of iron-rich silicates  is not considered in the present study because there are currently no established chemical nucleation routes that lead to the formation of such a material at high temperatures. The iron and magnesium abundances are very similar in the stellar photosphere, and iron seems not to be depleted in other dust types in the inner wind. Indeed, we do not form the molecules FeO and Fe$_2$ in significant amounts, two molecules that are essential to trigger the formation of iron oxides and pure iron grains. Most of the iron is in atomic form for $r < 10 $~\rs\ and might then easily be included in small silica-based clusters, as is the case for magnesium, to form Fe-rich silicates. If so, two silicate-based components are expected, one Mg-rich and the other Fe-rich, as was observed in laboratory study of a mixed Mg-Fe-SiO-H$_2$-O$_2$ vapour (\cite{riet99}). Because iron-rich silicate grains have high opacities around 1 \mic, they will be more efficient at driving a wind through radiation pressure and hotter than their iron-free counterparts, and will thus favour the emission of typical silicate features at 9.8 \mic\ and 18 \mic\ usually observed in the SEDs of Mira-type stars.

The inner wind gas conditions may be conducive to the synthesis of other solid compounds like spinel, \spin, and silica, \sili. Because of the efficient condensation of alumina very close to the star, and the availability of atomic magnesium (MgO does not form in large amounts at $r < 2$~\rs), the formation of spinel may occur along with that of alumina, or possibly compete with the latter. Furthermore, we have already mentioned that the production of alumina is triggered by the dimerisation of AlO. However, the large abundance of AlOH in the inner wind may induce the formation of molecular species not considered in the present study, such as HAlOH. Enhanced \al\ production may occur through HAlOH dimerisation and processes mimicking the proposed routes through silanone considered for silicate formation. Such chemical pathways have not yet been studied theoretically or experimentally, and we plan to investigate these new chemical routes and their impact on alumina production, combined with the formation of spinel and iron-rich silicates, in a future study. 

\begin{acknowledgement}
The authors thank the anonymous referee for comments that have improved the quality of the manuscript and Nicolas Mauron for stimulating discussions. 
\end{acknowledgement}

\end{document}